\documentclass[reprint,twocolumn,superscriptaddress,showkeys,
nofootinbib,notitlepage,amsmath,amssymb,floatfix]{revtex4-1}

\pdfoutput=1
\usepackage{latexsym,amsmath,amssymb,lmodern,float,url}
\usepackage{natbib}
\usepackage{color}
\usepackage{multirow}
\usepackage{bm}
\usepackage{array}

\def\de{\delta^{\vphantom{1}}}
\def\bde{{\bar\delta}}
\def\qq{{q\bar q}}
\def\QQ{{Q\bar Q}}

\def\bt{{\bar\theta}}

\def\h3{{\displaystyle{\frac 3 2}}}

\usepackage{tikz}
\usetikzlibrary{datavisualization}
\usetikzlibrary{datavisualization.formats.functions}
\usetikzlibrary{plotmarks} 
\usetikzlibrary{calc,arrows}
\newdimen\XCoord
\newdimen\YCoord

\begin{document}
\title{The Spectrum of $P$-Wave Hidden-Charm Exotic Mesons in the
Diquark Model}
\author{Jesse F. Giron}
\email{jfgiron@asu.edu}
\author{Richard F. Lebed}
\email{Richard.Lebed@asu.edu}
\affiliation{Department of Physics, Arizona State University, Tempe,
AZ 85287, USA}

\date{March, 2020}

\begin{abstract}
We study the fine structure in the spectrum of known and predicted
negative-parity hidden-charm exotic meson states, which comprise the
lowest $P$-wave multiplet in the dynamical diquark model.  Starting
with a form previously shown to successfully describe the $S$-wave
states, we develop a 5-parameter Hamiltonian that includes spin-orbit
and tensor terms.  After discussing the experimental status of the
observed $J^{PC} = 1^{--}$ states $Y$ with respect to masses and decay
modes (classified by eigenvalues of heavy-quark spin), we note a
number of inconsistencies between measurements from different
experiments that complicate a unique determination of the spectrum.
Outlining a variety of scenarios for interpreting the $Y$ data, we
perform fits to each one, obtaining results that demonstrate differing
possibilities for the $P$-wave spectra.  Choosing one of these fits
for illustration, we predict masses for all 28 isomultiplets in this
$1P$ multiplet, compare the results to tantalizing hints in the data,
and discuss the rich discovery potential for new states.
\end{abstract}

\maketitle

\section{Introduction}
\label{sec:Intro}

Although over 40 heavy-quark hidden-flavor exotic hadron candidates
have now been observed by various experiments in the past two decades,
no single theoretical picture to describe their structure has yet
emerged as the obvious favorite.  Several of the states [{\it e.g.},
$X(3872)$] lie provocatively close to 2-hadron thresholds and suggest
hadronic molecules or threshold effects, but others do not.  Some
prefer to decay only to conventional quarkonium states of one
particular heavy-quark spin eigenvalue ({\it e.g.}, $\psi$ rather than
$h_c$), hinting at particular quarkonium cores (as in the
hadroquarkonium model), but others [{\it e.g.}, $Z_b(10610) \! \to \!
\Upsilon, h_b$] mix heavy-quark spin eigenstates.   Some neutral
exotics [{\it e.g.}, $Y(4260)$] match the properties expected for
hybrid states.  And some [{\it e.g.}, $Z_c(4430)$] have thus far only
been clearly seen to decay to an excited quarkonium state [$\psi(2S)$]
even though the ground state ($J/\psi$) is available, suggesting a
large spatial extent for the exotic hadron wave function, which is a
feature of some diquark models.  The current experimental status of
the exotic hadron candidates, as well as detailed descriptions of the
aforementioned theoretical pictures, are presented in a number of
recent reviews~\cite{Lebed:2016hpi,Chen:2016qju, Hosaka:2016pey,
Esposito:2016noz,Guo:2017jvc,Ali:2017jda,Olsen:2017bmm,
Karliner:2017qhf,Yuan:2018inv,Liu:2019zoy,Brambilla:2019esw}.

In this paper we continue the development of the dynamical diquark
model, specifically by studying the spectroscopy of exotic
hidden-charm mesons of negative parity, which lie several hundred MeV
above the lowest (positive-parity) states.  The dynamical diquark
model is based upon a physical picture~\cite{Brodsky:2014xia} in which
4-quark states can be clearly identified as having a diquark $\de$ [in
the color-$\bar {\bf 3}$ combination $(Qq)_{\bar {\bf
3}}$]-antidiquark $\bde$ [$(\bar Q \bar q)_{\bf 3}$] structure only if
the $\de$-$\bde$ pair can achieve sufficient spatial separation as to
evade instantaneous recombination into 2-meson pairs $(Q\bar
q)$-$(\bar Q q)$ or $(\QQ)$-$(\qq)$.  The heavy-quark pair $\QQ$ in
this paper is exclusively $c\bar c$, although $b\bar b$ states can
also be analyzed this way, as well as $b\bar c$ states (if the full
state's discrete quantum numbers are handled
carefully~\cite{Lebed:2017min}), and also pentaquark
states~\cite{Lebed:2015tna} by using successive color-triplet
attractions to form a {\em triquark\/} $\bt \! \equiv \!
\left[ \bar Q (q_1 q_2)_{\bar {\bm 3}} \right]_{\bm 3}$ quasiparticle,
which can then form pentaquark states via the combination $\bt \de$.
Alternate constructions of triquark-diquark pentaquarks are studied in
Refs.~\cite{Karliner:2006hf,Ali:2019clg}.

This initial picture has been developed in stages into the dynamical
diquark model.  First, the nature of the state as a well-separated
$\de$-$\bde$ pair suggests a color flux tube connecting the pair,
meaning that the most natural formalism for describing their
spectroscopy uses multiplets of the Born-Oppenheimer (BO)
approximation, the same way that hybrid mesons have been treated for
decades~\cite{Griffiths:1983ah}.  The spectroscopy for $\de$-$\bde$
(and $\bt$-$\de$) states is presented in Ref.~\cite{Lebed:2017min}
(including detailed definitions of the standard BO potential
nomenclature, {\it e.g.}, $\Sigma^-_u$, {\it etc.}), and the most
relevant expressions are reprised in Sec.~\ref{sec:Spectrum} below;
one key result informing this work is that the lowest
[$\Sigma^+_g(1S)$] multiplet consists entirely of positive-parity
states, while the first excited multiplet [$\Sigma^+_g(1P)$] contains
only negative-parity states, a fact mentioned above.

The next step in building the model is to obtain explicit forms for
the BO potentials provided by the flux tube, which is accomplished by
numerically computing (primarily) gluonic potentials between static
heavy quarks on the lattice, as in Refs.~\cite{Juge:1997nc,
Juge:1999ie,Juge:2002br,Capitani:2018rox,Bali:2003jq}.  These
potentials are then fed into coupled Schr\"{o}dinger equations, which
are numerically solved to obtain the mass eigenvalues for the
$\de$-$\bde$ states.  The results of these
calculations~\cite{Giron:2019bcs} represent the spin- and
isospin-averaged masses for each multiplet, with the lowest few being
$1S$, $1P$, $2S$, $1D$, and $2P$, all in the $\Sigma^+_g$ potential.

In order to compare to actual measured exotic masses, one must
introduce spin and isospin splitting ({\it i.e.}, fine structure) into
the exotic multiplets by identifying significant physical effects
expected to break the mass degeneracy of the multiplet, such as
spin-spin couplings and spin-isospin dependent operators analogous to
those arising from pion exchange.  The relevant operators are
collected into a Hamiltonian, whose matrix elements embody the level
spacing of the mass spectrum.  The $1S$ multiplet is studied this way
in Ref.~\cite{Giron:2019cfc}, in which it is found that the essential
known phenomenological features of the lightest hidden-charm exotics
$X(3872)$, $Z_c(3900)$, and $Z_c(4020)$ emerge naturally.  These
techniques can also be applied directly to the $c\bar{c}s\bar{s}$
sector, as presented in a forthcoming study~\cite{Giron:ccss}.

Here we extend this study to the $1P$ multiplet, which as enumerated
in Ref.~\cite{Lebed:2017min} and again in Table~\ref{tab:PwaveStates}
below, contains 28 isomultiplets.  As to be discussed in
Sec.~\ref{sec:ExptReview}, no more than 6 of these states have been
observed, and indeed, the experimental situation for even these few
states remains unsettled.  Nevertheless, subsequent analysis by BESIII
and future measurements at Belle~II and elsewhere can be expected to
greatly clarify this murky picture.  In fact, by minimally extending
the Hamiltonian of Ref.~\cite{Giron:2019cfc} to include spin-orbit and
tensor terms, we provide a large number of testable predictions for
the unknown sectors of the multiplet, and with this explicit
Hamiltonian in hand, one can adjust detailed fits easily to
incorporate any future experimental modifications to the known
spectrum.  A study of the $P$-wave $\de$-$\bde$ states using QCD sum
rules appears in Refs.~\cite{Wang:2018ntv,Wang:2018ejf}.

This paper is organized as follows.  Section~\ref{sec:Spectrum}
summarizes the expected spectroscopy of the dynamical diquark model,
leading to a detailed enumeration of all 28 isomultiplets in the $1P$
multiplet.  In Sec.~\ref{sec:ExptReview} we review the current status
of exotic meson candidates in the mass range of 4050--4400~MeV ({\it
i.e.}, above the $1S$ states), identifying known $P \! = \! -$ states.
Special attention is given to ambiguities in the data for even the
best-known $J^{PC} \! = \! 1^{--}$ $Y$ states.  We then turn in
Sec.~\ref{sec:MassOps} to the development of a Hamiltonian for the
$1P$ states, starting with the form shown to be successful for
describing the phenomenology of the $1S$ states in
Ref.~\cite{Giron:2019cfc}.  Section~\ref{sec:Analysis} exhibits fits
of the data to experiment, using several hypotheses for identifying
the 4 $I \! = \! 0$, $1^{--}$ states of the $1P$ multiplet with the
observed $Y$ states.  We summarize briefly in Sec.~\ref{sec:Concl}.

\section{Spectroscopy of $P$-Wave $Q\bar Q q\bar q^\prime$ Exotics}
\label{sec:Spectrum}

The full spectroscopy of diquark-antidiquark ($\de$-$\bde$) exotics
connected by a gluonic field of arbitrary excitation quantum numbers,
and for arbitrary orbital excitations between the $\de$-$\bde$ pair,
is presented in Ref.~\cite{Lebed:2017min}, and much of the content in
this section mirrors Secs.~III and IV from that work.  As discussed in
Ref.~\cite{Lebed:2017min}, the excitations of the gluonic field
produce states analogous to the hybrids of ordinary quarkonium, and
may be classified according to the quantum numbers provided by
Born-Oppenheimer (BO) potentials.  However, detailed numerical
studies~\cite{Giron:2019bcs} show that such states lie above the
corresponding BO ground-state potential (called $\Sigma^+_g$) by at
least 1~GeV, comparable to the amount by which heavy-quarkonium
hybrids are expected from lattice simulations to lie above ordinary
quarkonium (see, {\it e.g.}, Ref.~\cite{Liu:2012ze}).  Since the
entire range of observed hidden-charm exotic candidates spans only
about 800~MeV~\cite{Lebed:2016hpi}, likely all of these states occupy
the $\Sigma^+_g$ BO potential, and specifically can be all
accommodated by the lowest $\Sigma^+_g$ levels: $1S$, $1P$, $2S$,
$1D$, and $2P$~\cite{Giron:2019bcs}.

A detailed accounting of the possible $Q\bar Q q\bar q^\prime$ states
(where the light quarks $q$ and $q^\prime$ do not necessarily carry
the same flavor) is straightforward for the $S$ wave, in which the
components possess no relative orbital angular momentum.  In this
case, any two naming conventions for the states differ only by the
order in which the 4 quark spins are coupled.  In the diquark basis,
in which the quark spins are coupled in the order $(qQ) + (\bar q \bar
Q)$, the 6 possible states are denoted by~\cite{Maiani:2014aja}:
\begin{eqnarray}
J^{PC} = 0^{++}: & \ & X_0 = \left| 0_\de , 0_\bde \right>_0 \, , \ \
X_0^\prime = \left| 1_\de , 1_\bde \right>_0 \, , \nonumber \\
J^{PC} = 1^{++}: & \ & X_1 = \frac{1}{\sqrt 2} \left( \left| 1_\de ,
0_\bde \right>_1 \! + \left| 0_\de , 1_\bde \right>_1 \right) \, ,
\nonumber \\
J^{PC} = 1^{+-}: & \ & Z \  = \frac{1}{\sqrt 2} \left( \left| 1_\de ,
0_\bde \right>_1 \! - \left| 0_\de , 1_\bde \right>_1 \right) \, ,
\nonumber \\
& \ & Z^\prime \, = \left| 1_\de , 1_\bde \right>_1 \, ,
\nonumber \\
J^{PC} = 2^{++}: & \ & X_2 = \left| 1_\de , 1_\bde \right>_2 \, ,
\label{eq:Swavediquark}
\end{eqnarray}
where outer subscripts indicate total quark spin $S$.  But the same
states may be expressed in any other basis by using angular momentum
recoupling coefficients; because one has a total of 4 angular momenta,
the $9j$ symbol applies.  For the purposes of this work, the most
useful alternate basis is that of definite heavy-quark (and
light-quark) spin, $(\QQ) + (\qq)$:
\begin{eqnarray}
\lefteqn{\left< (s_q \, s_{\bar q}) s_\qq , (s_Q \, s_{\bar Q}) s_\QQ
, S \, \right| \left. (s_q \, s_Q) s_\de , (s_{\bar q} \, s_{\bar Q})
s_\bde , S \right> } & & \nonumber \\
& = & \left( [s_\qq] [s_\QQ] [s_\de] [s_\bde] \right)^{1/2}
\left\{ \begin{array}{ccc} s_q & s_{\bar q} & s_\qq \\
s_Q & s_{\bar Q} & s_\QQ \\ s_\de & s_\bde & S \end{array} \! \right\}
\, , \ \ \label{eq:9jTetra}
\end{eqnarray}
where $[s] \! \equiv \! 2s \! + \! 1$ denotes the multiplicity of a
spin-$s$ state.  Using Eqs.~(\ref{eq:Swavediquark}) and
(\ref{eq:9jTetra}), one then obtains
\begin{eqnarray}
J^{PC} = 0^{++}: & \ & X_0 \equiv \frac{1}{2} \left| 0_\qq , 0_\QQ
\right>_0 + \frac{\sqrt{3}}{2} \left| 1_\qq , 1_\QQ \right>_0 \, ,
\nonumber \\
& & X_0^\prime \equiv \frac{\sqrt{3}}{2} \left| 0_\qq , 0_\QQ
\right>_0 - \frac{1}{2} \left| 1_\qq , 1_\QQ \right>_0 \, , 
\nonumber \\
J^{PC} = 1^{++}: & \ & X_1 \equiv \left| 1_\qq , 1_\QQ \right>_1 \, ,
\nonumber \\
J^{PC} = 1^{+-}: & \ & Z \; \equiv \frac{1}{\sqrt 2} \left( \left| 
1_\qq , 0_\QQ \right>_1 \! - \left| 0_\qq , 1_\QQ \right>_1 \right)
\, , \nonumber \\
& \ & Z^\prime \equiv \frac{1}{\sqrt 2} \left( \left| 1_\qq ,
0_\QQ \right>_1 \! + \left| 0_\qq , 1_\QQ \right>_1 \right) \, ,
\nonumber \\
J^{PC} = 2^{++}: & \ & X_2 \equiv \left| 1_\qq , 1_\QQ \right>_2 \, .
\label{eq:SwaveQQ}
\end{eqnarray}
A similar recoupling can be used to express these states in terms of
equivalent heavy-light meson spins, $(q \bar Q) + (\bar q Q)$.

The pairs of states $X_0, X^\prime_0$, and $Z, Z^\prime$ carry the
same value of $J^{PC}$ (their discrete quantum numbers to be
discussed below), and can therefore mix.  One may define the
equivalent unique heavy-quark spin eigenstates, which are: $X_1$,
$X_2$, and
\begin{eqnarray}
{\tilde X}_0 & \equiv & \left| 0_\qq , 0_\QQ \right>_0 =
+ \frac{1}{2} X_0 + \frac{\sqrt{3}}{2} X_0^\prime \, , \nonumber \\
{\tilde X}_0^\prime & \equiv & \left| 1_\qq , 1_\QQ \right>_0 =
+ \frac{\sqrt{3}}{2} X_0 - \frac{1}{2} X_0^\prime \, , \nonumber \\
{\tilde Z} & \equiv & \left| 1_\qq , 0_\QQ \right>_1 =
\frac{1}{\sqrt{2}} \left( Z^\prime \! + Z \right) \, , \nonumber \\
{\tilde Z}^\prime & \equiv & \left| 0_\qq , 1_\QQ \right>_1 =
\frac{1}{\sqrt{2}} \left( Z^\prime \! - Z \right) \, .
\label{eq:HQbasis}
\end{eqnarray}
Up to this point, the light quarks $q$ and $\bar q$ have been taken to
carry the same flavor.  Allowing for a potentially distinct light
flavor $\bar q^\prime$ for the antiquark then introduces the isospin
quantum number into the analysis.\footnote{If the $s$ quark is also
considered light, one may generalize to consider states in SU(3)$_{\rm
flavor}$ multiplets.}  Each state in Eqs.~(\ref{eq:Swavediquark}) or
(\ref{eq:HQbasis}) then occurs in $I \! = \!  0$ and $I \! = \! 1$
variants, doubling the list to 12 possible isomultiplets.
Experimentally, both $I \! = \! 0$ and $I \! = \! 1$ exotic candidates
have been observed, making the inclusion of isospin an essential
ingredient for an accurate phenomenological analysis.  In the
nomenclature adopted by the Particle Data Group
(PDG)~\cite{Tanabashi:2018oca}, the observed $S$-wave states
$X_0$/$X^\prime_0$, $X_1$, $Z$/$Z^\prime$, and $X_2$ with $I \! = \!
0$ are to be denoted as $\chi_{c0}$, $\chi_{c1}$, $h_c$, and
$\chi_{c2}$, respectively, while those with $I \! = \! 1$ are to be
denoted as $W_{c0}$, $W_{c1}$, $Z_c$, and $W_{c2}$, respectively.

Allowing now for an orbital excitation quantum number $L$ between the
$\de$-$\bde$ pair (but not within either of $\de$ or $\bde$), the
parity eigenvalue $P$ is simply the usual spatial factor $(-1)^L$,
since the intrinsic parity factor $-1$ attached to an antiquark
appears twice.  One immediately notes that all 12 states in the $1S$
band have $P \! = \! +$; and in the hidden-charm sector, this feature
is in accord with experiment.  Namely, all exotic candidates supported
by strong experimental evidence and with known parity, stretching from
the $X(3872)$ up to a mass of 4200~MeV, have $P \! = \! +$ and fit
nicely into the $1S$ level.  In fact, the explicit numerical
calculations of Ref.~\cite{Giron:2019bcs} show that the centroids of
the $1S$ and $1P$ multiplets are separated by about 370~MeV.
Meanwhile, Ref.~\cite{Giron:2019cfc} presents an analysis of mass
splittings among the 12 hidden-charm $1S$ states (both known and
predicted), and shows that the $X(3872)$ can emerge naturally as the
lightest member of the multiplet.  Building upon this analysis, the
current paper focuses upon the $1P$ states, all of which carry $P \! =
\! -$.

The charge-conjugation quantum number $C$ is most easily determined in
the $(\QQ) + (\qq)$ basis, since the exchange of each quark with its
antiquark requires both a spatial inversion factor $(-1)^L$
(exchanging $\de \! \leftrightarrow \! \bde$) and spin-exchange
factors $(-1)^{s_\QQ + 1}$ and $(-1)^{s_\qq + 1}$.  Thus $C \! = \!
(-1)^{L + s_\QQ + s_\qq}$, which for $L \! = \! 0$ gives the $C$
eigenvalues listed in Eqs.~(\ref{eq:HQbasis}).  The same analysis
holds if the state may be considered in a molecular picture, for which
the orbital excitation $L$ connects the heavy-light components $(q
\bar Q) + (\bar q Q)$.

Lastly, $C$ is a good quantum number only for states with $q^\prime \!
= \! q$, {\it i.e.}, the $I_3 \! = \! 0$ (neutral) members of
isomultiplets.  More generally, the good quantum number relevant for
both neutral and charged states is $G$ parity, $G \! = \! (-1)^I C$,
where $C$ refers to the neutral member of the isomultiplet.  In
summary, one has
\begin{eqnarray}
P & = & (-1)^L \, , \nonumber \\
C & = & (-1)^{L + s_\QQ + s_\qq} \ \ {\rm (neutral \ states)} \, ,
\nonumber \\
G & = & (-1)^{I + L + s_\QQ + s_\qq} \, . \label{eq:Discrete}
\end{eqnarray}

Considering, finally, the $P$-wave states that form the subject of
this paper, one enumerates them\footnote{The number of $P$-wave
$\de$-$\bde$ states was first counted in Ref.~\cite{Cleven:2015era},
and the corresponding counting for all partial waves appears in
Ref.~\cite{Lebed:2017min}.} simply by combining the $S$-wave ``core''
states of Eqs.~(\ref{eq:Swavediquark}) or (\ref{eq:HQbasis})
(generically denoted here by $Y$) with a unit of orbital angular
momentum, $L \! = \! 1$, and imposes the usual rules of angular
momentum addition to form eigenstates of total $J$.  The notation
introduced in Ref.~\cite{Lebed:2017min} then reads $Y^{(J)}_L$; since
we are interested only in $P$-wave states, the subscript $L \! = \! 1$
is suppressed.  Using the heavy-quark basis of
Eqs.~(\ref{eq:HQbasis}), one obtains the 28 isomultiplets listed in
Table~\ref{tab:PwaveStates}, 14 with $I \! = \! 0$ and 14 with $I \!
= \!  1$.
\begin{table}[ht]
\caption{The 28 $c\bar c q \bar q^\prime$ isomultiplets for the 
$\Sigma_g^+\left(1P\right)$ band in the
dynamical diquark model.  The given state notation, using the basis of
Eqs.~(\ref{eq:HQbasis}), uniquely specifies eigenvalues of light-quark
spin $s_\qq$, heavy-quark spin $s_\QQ$, total quark spin $S$, and
total angular momentum $J$.  Discrete quantum numbers $P$, $C$ are
obtained from Eqs.~(\ref{eq:Discrete}).  Names of $I \! = \! 0$ and $I
\! = \! 1$ states according to Particle Data Group
conventions~\cite{Tanabashi:2018oca} are also presented.  $J_\qq$
[Eq.~(\ref{eq:JqqDef})] denotes eigenvalues of the total angular
momentum carried by light quarks, and $\mathcal{M}_{J_\qq}$
[Eq.~(\ref{eq:JqqCoef})] gives the amplitude of each allowed $J_\qq$
eigenstate within the given state.}
\label{tab:PwaveStates}\centering
\setlength{\extrarowheight}{1.2ex}
\begin{tabular}{ccccccccc}
\hline\hline
State & $J^{PC}$ & $I=0$ & $I=1$ & $s_\qq$ & $s_\QQ$ & $S$ & $J_\qq$ &
$\mathcal{M}_{J_\qq}$ \\[0.5ex]
\hline
$\tilde{X}_0^{(1)}$ & $ 1^{--}$ & $\psi$ & $R_{c1}$ & $0$ & $0$ & $0$ & $1$ & $+1$ \\
$\tilde{X}_0'^{(1)}$ & $ 1^{--}$ & $\psi$ & $R_{c1}$ & $1$ & $1$ & $0$ & $0$ & $+\frac{1}{3}$\\
$ $ & $ $ & $ $ & $ $ & $ $ & $ $ & & $1$ & $-\frac{1}{\sqrt{3}}$\\
& $ $ & $ $ & $ $ & $ $ & $ $ & $ $ & $2$ & $+\frac{\sqrt{5}}{3}$\\
$X_1^{(1)}$ & $ 1^{--}$ & $\psi$ & $R_{c1}$ & $1$ & $1$ &$1$ & $0$ & $-\frac{1}{\sqrt{3}}$\\
& $ $ & $ $ & $ $ & $ $ & $ $ & $ $ & $1$ & $+\frac{1}{2}$\\
& $ $ & $ $ & $ $ & $ $ & $ $ & $ $ & $2$ &$+\frac{1}{2}\sqrt{\frac{5}{3}}$\\
$X_2^{(1)}$ & $ 1^{--}$ & $\psi$ & $R_{c1}$ & $1$ & $1$ & $2$ & $0$ & $+\frac{\sqrt{5}}{3}$\\
& $ $ & $ $ & $ $ & $ $ & $ $ & $ $ & $1$ & $+\frac{1}{2}\sqrt{\frac{5}{3}}$\\
& $ $ & $ $ & $ $ & $ $ & $ $ & $ $ & $2$ &$+\frac{1}{6}$\\
$X_1^{(0)}$ & $ 0^{--}$ & $\psi_0$ & $R_{c0}$ & $1$ & $1$ & $1$ & $1$ & $+1$\\
$X_1^{(2)}$ & $ 2^{--}$ & $\psi_2$ & $R_{c2}$ & $1$ & $1$ & $1$ & $1$ & $-\frac{1}{2}$\\
& $ $ & $ $ & $ $ & $ $ & $ $ & $ $ & $2$ & $+\frac{\sqrt{3}}{2}$\\
$X_2^{(2)}$ & $ 2^{--}$ & $\psi_2$ & $R_{c2}$ & $1$ & $1$ & $2$ & $1$ & $+\frac{\sqrt{3}}{2}$\\
& $ $ & $ $ & $ $ & $ $ & $ $ & $ $ & $2$ & $+\frac{1}{2}$\\
$X_2^{(3)}$ & $ 3^{--}$ & $\psi_3$ & $R_{c3}$ & $1$ & $1$ & $2$ & $2$ & $+1$\\
$\tilde{Z}^{(0)}$ & $ 0^{-+}$ & $\eta_c$ & $\Pi_{c0}$ & $1$ & $0$ & $1$ & $0$ & $+1$\\
$\tilde{Z}'^{(0)}$ & $ 0^{-+}$ & $\eta_c$ & $\Pi_{c0}$ & $0$ & $1$ & $1$ & $1$ & $+1$\\
$\tilde{Z}^{(1)}$ & $ 1^{-+}$ & $\eta_{c1}$ & $\Pi_{c1}$ & $1$ & $0$ & $1$ & $1$ & $+1$\\
$\tilde{Z}'^{(1)}$ & $ 1^{-+}$ & $\eta_{c1}$ & $\Pi_{c1}$ & $0$ & $1$ & $1$ & $1$ & $+1$\\
$\tilde{Z}^{(2)}$ & $ 2^{-+}$ & $\eta_{c2}$ & $\Pi_{c2}$ & $1$ & $0$ & $1$ & $2$ & $+1$\\
$\tilde{Z}'^{(2)}$ & $ 2^{-+}$ & $\eta_{c2}$ & $\Pi_{c2}$ & $0$ & $1$ & $1$ & $1$ & $+1$\\
\hline\hline
\end{tabular}
\end{table}

\section{Experimental Review of Relevant States}
\label{sec:ExptReview}

The hidden-charm exotic-meson candidates with masses above the
well-known $X(3872)$, $Z_c(3900)$, and $Z_c(4020)$, but below that of
the $Z_c(4430)$, are listed in Table~\ref{tab:Expt}.  This range has
been chosen in order to eliminate many states clearly established to
carry positive parity (such as those just listed), but also to bracket
the well-known $J^{PC} \!  = \! 1^{--}$ states $Y(4230)$, $Y(4260)$,
$Y(4390)$, and $Y(4360)$.  The higher-mass $1^{--}$
$Y(4626)$~\cite{Jia:2019gfe,Jia:2020epr} and $Y(4660)$ are also
included, being the only hidden-charm exotic candidates with clearly
established negative parity that lie outside this range.

The only other observed $P \! = \! -$ exotic candidate, the
(unconfirmed) $Z_c(4240)$, carries quark-model exotic quantum numbers
$J^{PC} \! = \! 0^{--}$ as well as $I \! = \! 1$, and also lies in the
range 4020--4480~MeV\@.  The only states inside this range confirmed
to carry $P \! = \! +$ are $\chi_{c1}(4140)$ and $\chi_{c1}(4274)$
(both $1^{++}$), and $Z_c(4200)$ ($1^{+-}$).  However, the former 2
states have only been seen in decays to $J/\psi \, \phi$, and may
therefore be $c\bar c s\bar s$ states~\cite{Lebed:2016yvr} (and
similarly for the $P \! = \! -$ $Y(4626)$ due to its $D_s^+
D_{s1,s2}^-$ decays~\cite{Jia:2019gfe,Jia:2020epr}), while the
existence of $Z_c(4200)$ remains unconfirmed.

\begin{table*}[ht]
  \caption{All charmoniumlike exotic-meson candidates catalogued by
  the Particle Data Group (PDG)~\cite{Tanabashi:2018oca} to lie in
  mass above $Z_c(4020)$ and below $Z_c(4430)$. Also included are
  $Y(4626)$ and $Y(4660)$, the only other charmoniumlike exotic mesons
  observed to carry $P \! = \! -$.  Both the particle name most
  commonly used in the literature and its label as given in the PDG
  are listed.}
\label{tab:Expt}
\centering
\setlength{\extrarowheight}{1.2ex}
\begin{tabular}{cccccc}
\hline\hline
    Particle
        & PDG label
            & $I^{G}J^{PC}$
                & Mass [MeV]
                    & Width [MeV]
                        & Production and decay \\ \hline
    $Z_c(4050)$ & $X(4050)^\pm$
        & $1^-?^{?+}$
            & $4051^{+24}_{-40}$
                & $82^{+50}_{-28}$
                    & $B \to KZ$; $Z \to \pi^\pm \chi_{c1}$\\ %
    $Z_c(4055)$ & $X(4055)^\pm$
        & $1^+?^{?-}$
            & $4054 \pm 3.2$
                & $ 45 \pm 13$
                    & $e^+e^- \to \gamma \pi^\mp Z$; $Z \to \pi^\pm \psi(2S)$ \\ %
    $Z_c(4100)$ & $X(4100)^\pm$
        & $1^- ?^{??}$
            & $4096 \pm 28$
                 & $152^{+80}_{-70}$
                    & $B \to K Z; Z \to  \pi^\pm \eta_c$ \\
    $Y(4140)$ & $\chi_{c1}(4140)$
        & $0^+1^{++}$
            & $4146.8 \pm 2.4$
                & $22^{+8}_{-7}$
                    &  $\begin{array}{r} B \to K Y \\
		 p\bar{p} \to Y + {\rm anything} \end{array} \bigg\}$ ;
                      $Y \to \phi J/\psi$ \\ %
    $X(4160)$ & $X(4160)$
        & $?^? ?^{??}$
            & $4156^{+29}_{-25}$
                & $139^{+110}_{-60}$
                    & $e^+e^- \to J/\psi + X$; $X \to D^* {\bar D}^*$ \\ %
    $Z_c(4200)$ & $Z_c(4200)$
        & $1^+1^{+-}$
            & $4196^{+35}_{-32}$
                & $370^{+100}_{-150}$
                    & $B \to KZ$; $Z \to \pi^\pm J/\psi$ \\ %
    $Y(4230)$ & $\psi(4230)$
        & $0^-1^{--}$
            & $4218^{+5}_{-4}$
                & $59^{+12}_{-10}$
                    & $e^+e^- \to Y$; $Y \to \left\{ \begin{array}{l}
                     \omega \chi_{c0} \\
                     \eta J/\psi\\
                     \pi^+ \pi^- h_c \\
                     \pi^+ \pi^- \psi(2S) \\
		     \pi^+ D^0 D^{*-} \\
		\end{array} \right. $ \\ %
    $Z_c(4240)$ & $R_{c0}(4240)$
        & $1^+0^{--}$
            & $4239^{+50}_{-21}$
                & $220^{+120}_{-90}$
                    & $B\to KZ$; $Z \to \pi^\pm \psi(2S)$ \\ %
    $Z_c(4250)$ & $X(4250)^\pm$
        & $1^-?^{?+}$
            & $4248^{+190}_{-50}$
                & $177^{+320}_{-70}$
                    & $B \to KZ$; $Z \to \pi^\pm \chi_{c1}$ \\ %
    $Y(4260)$ & $\psi(4260)$
        & $0^-1^{--}$
            & $4230 \pm 8$
                & $55 \pm 19$
                    & $e^+e^- \to \gamma Y$ or $Y$; $Y \to \left\{ \begin{array}{l}
                        \pi^+ \pi^- J/\psi \\
                        f_0(980) J/\psi \\
                       \pi^\mp Z_c^\pm (3900) \\
                       K^+ K^- J/\psi \end{array} \right. $ \\ %
    $Y(4274)$ & $\chi_{c1}(4274)$
        & $0^+1^{++}$
            & $4274^{+8}_{-6}$
                & $49 \pm 12$
                    & $B \to K Y$; $Y \to \phi J/\psi$ \\ %
    $X(4350)$ & $X(4350)$
        & $0^+?^{?+}$
            & $4351 \pm 5$
                & $ 13^{+18}_{-10}$
                    & $\gamma\gamma \to X$; $X\to \phi J/\psi$ \\ %
    $Y(4360)$ & $\psi(4360)$
        & $0^- 1^{--}$
            & $4368 \pm 13$
                & $96 \pm 7$
                    & $e^+e^- \to \gamma Y$ or $Y$; $Y \to \bigg\{ \begin{array}{l}
                        \pi^+ \pi^- \psi(2S) \\
                        \pi^0 \pi^0 \psi(2S) \end{array}$ \\ %
    $Y(4390)$ & $\psi(4390)$
        & $0^- 1^{--}$
            & $4392 \pm 7$
                & $140^{+16}_{-21}$
                    & $e^+e^- \to Y$; $Y \to \bigg\{ \begin{array}{l}
			\eta J/\psi \\		        
			\pi^+ \pi^- h_c \end{array}$ \\ %
    $Y(4626)$ & $\psi(4626)$
        & $0^- 1^{--}$
            & $4626 \pm 6$
                & $49^{+14}_{-12}$
                    & $e^+e^- \to \gamma Y$; $Y \to D_s^+
			D_{s1}(2536)^- [{\rm or} \ D_s^+
			D^*_{s2}(2573)^-]$ \\ %
    $Y(4660)$ & $\psi(4660)$
        & $0^- 1^{--}$
            & $4643 \pm 9$
                & $72 \pm 11$
                    & $e^+e^- \to \left\{ \begin{array}{l}
                        \gamma Y; Y \to \pi^+ \pi^- \psi(2S) \\
                        \ \; Y; Y \to \Lambda^+_c \Lambda^-_c
                        \end{array} \right.$ \\ \hline\hline
\end{tabular}
\end{table*}

The currently known data set of $P \! = \! -$ exotics therefore
consists of the 6 states $Y(4230)$, $Z_c(4240)$, $Y(4260)$, $Y(4360)$,
$Y(4390)$, $Y(4626)$, and $Y(4660)$.  $Z_c(4240)$ has only been
observed in the LHCb experiment that confirmed the existence of the
$1^{+-}$ $Z_c(4430)$~\cite{Aaij:2014jqa}, and both states are clearly
$I \! = \! 1$ because they are seen in decays to $\psi(2S) \,
\pi^\pm$.  Yet even the well-studied $Y$ states present a number of
mysteries.  Each one has only been produced either via initial-state
radiation processes ($e^+ e^- \! \to \! \gamma Y$) at BaBar, CLEO, and
Belle, or directly via $e^+ e^- \! \to \! Y$ at BESIII\@.  Unlike the
conventional charmonium $1^{--}$ states, none of them produce obvious
peaks in $R(e^+ e^- \! \to \!  {\rm hadrons})$ (see Fig.~51.3 in
Ref.~\cite{Tanabashi:2018oca}).  Their only open-charm decays yet
observed are $Y(4230) \! \to \! \pi^+ D^0
D^{*-}$~\cite{Ablikim:2018vxx}, $Y(4626) \! \to \! D_s^+
D_{s1,s2}^-$~\cite{Jia:2019gfe,Jia:2020epr}, and $Y(4660) \! \to
\Lambda^+_c \Lambda^-_c$, and this latter mode has not been confirmed
to belong specifically to $Y(4660)$.  $Y(4230)$, $Y(4260)$, $Y(4360)$,
and $Y(4626)$ are confirmed to be $I \! = \! 0$ due to their observed
decays to $\chi_{c0} \, \omega$, $J/\psi \, f_0(980)$, $\psi(2S) \,
\pi^0 \pi^0$, and $D_s D_{s1,s2}$, respectively.  But $Y(4390)$ and
$Y(4660)$ (not counting the latter state's possible $\Lambda^+_c
\Lambda^-_c$ mode) do not yet have confirmed isospin quantum
numbers, and could easily carry $I \! = \! 1$, due to their production
via a virtual photon.

The pattern of observed $Y$ decays to conventional charmonium is
itself quite confusing: Referring to Table~\ref{tab:Expt}, the
$Y(4230)$ has been seen to decay to $\psi(2S)$, $\chi_{c0}$, and
$h_c$, while $Y(4260)$ is so far only observed in channels decaying
(eventually) to $J/\psi$; $Y(4360)$ to both $J/\psi$ and $\psi(2S)$;
$Y(4660)$ only to $\psi(2S)$; and $Y(4390)$ only to $h_c$.
Recent data from BESIII~\cite{ablikim2020observation} show that
$Y(4230)$ and $Y(4390)$ both decay to $J/\psi \, \eta$, which restricts 
$Y(4390)$ to $I \! = \! 0$.

However, closer inspection of the data introduces yet another level of
ambiguity.  BESIII, having the world's most complete data set on
$1^{--}$ states, does not confirm the location of $Y(4260)$ given by
earlier experiments (which give well-clustered average values $m \! =
\! 4251 \! \pm \! 6$~MeV, $\Gamma \! = \! 120 \! \pm \!  12$~MeV),
and instead reports $m \! = \! 4222.0 \! \pm \! 3.4$~MeV and $\Gamma
\! = \! 44.1 \! \pm \! 4.7$~MeV~\cite{Ablikim:2016qzw}.  Note that the
former numbers appear in $e^+ e^- \! \to \! \gamma Y$ measurements and
the latter directly from $e^+ e^- \! \to \! Y$.  Indeed, the BESIII
numbers for $Y(4260)$ actually agree rather well with those given in
Table~\ref{tab:Expt} for $Y(4230)$, which is especially true now that
the current values extracted from $\chi_{c0} \, \omega$ have been
updated~\cite{Ablikim:2019apl}:  $m \! = \! 4218.5 \! \pm \! 4.3$~MeV
and $\Gamma \! = \! 28.2 \! \pm \! 4.2$~MeV.  Similarly, the BESIII
mass measurements ascribed to $Y(4360)$ give two widely separated
values, $4320 \! \pm \! 13$~MeV for $e^+ e^- \! \to \! J/\psi \,
\pi^+ \pi^- $~\cite{Ablikim:2016qzw} versus $4386 \! \pm \! 4$~MeV for
$e^+ e^- \! \to \! \psi(2S) \pi^+ \pi^-$~\cite{Ablikim:2017oaf}.

It is worth noting, in such a fluid experimental landscape, that
signals for decays into particular channels currently ascribed to a
single resonance might actually represent observations of more than
one resonance closely spaced in mass.  For example, the isolated
low-mass measurement of $e^+ e^- \! \to \! J/\psi \, \pi^+ \pi^- $
from Ref.~\cite{Ablikim:2016qzw} could easily signal a resonance
distinct from $Y(4360)$, and indeed, $I \! = \! 1$ cannot yet be ruled
out for this mode.  Until such time as this mass range is fully
explored in the charged sector,\footnote{For example, a very recent
experiment by LHCb~\cite{Aaij:2019ipm} shows resonant structure in
$J/\psi \, \pi^-$ around 4200 and around 4600~MeV.} a definitive
description of the splitting of quartets of states into $I \! = \! 0$
and $I \! = \! 1$ multiplets remains unavailable.

Lastly, we do not consider in this work the broad $1^{--}$ $J/\psi \,
\pi^+ \pi^-$ resonance $Y(4008)$ claimed by Belle~\cite{Yuan:2007sj},
as it has not been seen by other experiments, particularly in the
precise scan of BESIII~\cite{Ablikim:2016qzw}.  We also do not
consider the charged~\cite{Ablikim:2017oaf} and
neutral~\cite{Ablikim:2017aji} charmoniumlike $\psi(2S) \pi$
``structures'' observed by BESIII around 4035~MeV, which may be the
same as $Z_c(4055)$~\cite{Tanabashi:2018oca}.

\section{Mass Hamiltonian Operators}
\label{sec:MassOps}

\subsection{Operators Appearing for $1S$ and $1P$}

The analysis of the $1S$ states in Ref.~\cite{Giron:2019cfc} uses a
Hamiltonian consisting of only 3 operators,
\begin{eqnarray}
H & = & M_0 + 2 \kappa_{qQ} ({\bf s}_q \! \cdot \! {\bf s}_Q +
{\bf s}_{\bar q} \! \cdot \! {\bf s}_{\bar Q}) + V_0 \, {\bm \tau}_q
\! \cdot \! {\bm \tau}_{\bar q} \; {\bm \sigma}_q \! \cdot \!
{\bm \sigma}_{\bar q} \, , \hspace{1em}
\label{eq:Ham}
\end{eqnarray}
where $M_0$ is the common multiplet mass; the $\kappa_{qQ}$ term
indicates that the only isospin-blind spin couplings are taken to be
only those between pairs of quarks within the same diquark: $q \!
\leftrightarrow \! Q$ in $\de$ and $\bar q \! \leftrightarrow \!
\bar Q$ in $\bde$; and $V_0$ is a spin-isospin operator identical in
form to that appearing in the canonical one-pion exchange between
nucleons, except that it is taken to act only between the light quarks
$q, \bar q$ rather than between the full diquarks.  Isospin-blind
spin-spin couplings between quarks within different diquarks ({\it
e.g.}, $q$ and $\bar Q$) are ignored in Ref.~\cite{Giron:2019cfc}
under the assumption that $\de, \bde$ are somewhat separated
quasiparticles within the hadron, and thus isospin-blind interactions
between $\de$ and $\bde$ are assumed to be relatively small compared
to the spin couplings inside of them.  An alternative choice for the
dominant isospin-dependent operator is also considered in
Ref.~\cite{Giron:2019cfc},
\begin{equation}
\Delta H = V_1 \, {\bm \tau}_q \! \cdot \! {\bm \tau}_{\bar q} \;
{\bm \sigma}_\de \! \cdot \! {\bm \sigma}_\bde \, ,
\label{eq:IsospinDiquark}
\end{equation}
in which the isospin-dependent interaction is chosen to couple to the
full $\de$ or $\bde$ as a unit (noting that $Q, \bar Q$ are
isosinglets, so that ${\bm \tau}_\de \! = \! {\bm \tau}_q$).  However,
such an operator leaves the $J^{PC} \!  = \! 1^{++}$ states with $I \!
= \! 0$ and $I \! = \! 1$ in the $1S$ multiplet degenerate in mass.
Since one of the key experimental properties of the $1^{++}$ $X(3872)$
is its lack of charged partners~\cite{Choi:2011fc}, the operator in
Eq.~(\ref{eq:IsospinDiquark}) cannot give the dominant
isospin-dependent effect for the $1S$ states.  We inferred in
Ref.~\cite{Giron:2019cfc} that the isospin-dependent couplings, at
least in the $S$ wave, appear to see only the light quarks, and do not
view the diquarks as pointlike constituents.  Indeed, one of the major
thrusts of Ref.~\cite{Giron:2019cfc} is a study showing that the
spectrum of the diquark model is quite robust when the $\de,
\bde$ are endowed with wave functions of significant spatial
extent compared to the size of the full hadron and begin to overlap.
In such a situation, it is quite natural that isospin-dependent
couplings are sensitive only to the light quarks $q
\bar q^\prime$, independent of the heavy quarks $\QQ$.

The Hamiltonian of Eq.~(\ref{eq:Ham}) is shown in~\cite{Giron:2019cfc}
to accommodate the known $P \! = \! +$ exotics $X(3872)$, $Z_c(3900)$,
and $Z_c(4020)$, while predicting the masses of the other 9 yet-unseen
states in the $1S$ multiplet in such a way that the $X(3872)$ emerges
naturally as the lightest state, $Z_c(3900)$ preferentially decays to
$J/\psi \, \pi$, and $Z_c(4020)$ preferentially decays to $h_c \,
\pi$, in complete accord with experiment.  Nevertheless, the
Hamiltonian of Eq.~(\ref{eq:Ham}) contains only operators expected to
provide the most significant physical effects.  While it is possible
to fit the 3 constants $M_0$, $\kappa_{qQ}$, $V_0$ using nothing but
the 3 PDG-averaged mass eigenvalues of $X(3872)$, $Z_c(3900)$, and
$Z_c(4020)$, and indeed such a restrictive fit gives results virtually
identical to those in Ref.~\cite{Giron:2019cfc}:
\begin{eqnarray} \label{eq:LastPaper}
M_0 (1S)         & = & 3988.69 \ {\rm MeV} \, , \nonumber \\
\kappa_{qQ} (1S) & = &   17.89 \ {\rm MeV} \, , \nonumber \\
V_0 (1S)         & = &   33.04 \ {\rm MeV} \, ,
\end{eqnarray}
this fit ignores the effect of subleading operators that can have a
significant effect upon small mass splittings and mixing angles, as
well as the inevitable shifting of PDG central mass values expected to
occur as newer measurements of these masses from subsequent
experiments are published.  For example, $\kappa_{qQ}$ is particularly
sensitive to the $Z_c(3900)$ mass, and would more than double if its
mass measurement increased by only 10~MeV\@. In addition, the values
in Eqs.~(\ref{eq:LastPaper}) apply only to states in the $1S$
multiplet and must be re-evaluated for the $1P$ state, since they
represent expectation values over wave functions that vary for states
of different quantum numbers.

\subsection{Operators First Appearing for $1P$}

Turning now to the $1P$ states, the multiplet-average mass $M_0 (1P)$
can be obtained either by using mass eigenvalues of $P \! = \! -$
states alone, or by applying the simulations of
Ref.~\cite{Giron:2019bcs} and fixing as an initial point the value of
the multiplet-average mass $M_0(1S)$ from Eq.~(\ref{eq:LastPaper}).
Using the latter approach, and depending upon which specific lattice
simulation is used to obtain the $\Sigma^+_g$ BO potential, one
obtains
\begin{equation} \label{eq:Mass1P}
M_0 (1P) = 4358 \mbox{-} 4361 \ {\rm MeV} \, .
\end{equation}
In a purely phenomenological analysis~\cite{Maiani:2014aja} spanning
the two multiplets, the $1P$-$1S$ splitting is provided by the
operator ${\bf L}^2$.  The next most significant $L$-dependent
contribution is expected to arise from the spin-orbit term,
\begin{equation} \label{eq:SpinOrbit}
\Delta H_{LS} = V_{LS} \, {\bf L} \cdot {\bf S} \, ,
\end{equation}
noting that the analogous parameter $V_{LS}$ defined in
Ref.~\cite{Maiani:2014aja} is a factor of $-\frac 1 2$ as large as the
one defined in Eq.~(\ref{eq:SpinOrbit}).  Explicitly, its matrix
elements contribute to the mass an amount
\begin{equation}
\Delta M_{LS} = \frac{V_{LS}}{2} [J(J+1)-L(L+1)-S(S+1)] \, .
\end{equation}

The parameter $\kappa_{qQ}$ from Eq.~(\ref{eq:Ham}), providing matrix
elements contributing to the mass as
\begin{equation}
\Delta M_{\kappa_{qQ}} = \kappa_{qQ} \left[ s_\de (s_\de + 1) +
s_{\bde} (s_{\bde} + 1) - 3 \right] \, ,
\end{equation}
and representing spin-only couplings within a diquark, might be
expected to vary with the mass of the heavy quark $Q$,\footnote{In
this work $Q \! = \! c$, but the same set of operators (with different
numerical values for the coefficients) is expected to apply to
hidden-bottom and $B_c$-like exotic systems.} but inasmuch as the
diquarks $\de, \bde$ are compact and well separated within the
hadron---which is not such a clear-cut proposition, in light of the
discussion following Eq.~(\ref{eq:IsospinDiquark})---one would naively
expect $\kappa_{qQ}$ to be roughly the same for all multiplets of a
fixed flavor content.  However, as seen below, the best fits to
$\kappa_{qQ}$ in the $1P$ sector favor rather larger values.
Nevertheless, in light of both the sensitivity of $\kappa_{qQ}(1S)$ to
$m_{Z_c(3900)}$ noted above, and of questions about the detailed
consequences of finite diquark size, such larger values need not be
construed as problematic.

The parameter $V_0$ from Eq.~(\ref{eq:Ham}), providing contributions
to the mass of the form
\begin{equation}
\Delta M_{V_0} = V_0 \left[ 2I(I+1) - 3 \right]
\left[ 2s_{\qq} (s_{\qq} +1) - 3 \right] \, ,
\end{equation}
is introduced in Ref.~\cite{Giron:2019cfc} to represent a pionlike
coupling in the colored environment of the flux tube connecting the
$\de$-$\bde$ pair.  As is well known, the corresponding operator
representing pion exchange between two nucleons has a positive
coefficient, indicating attractive $I \! = \!  1$, spin-singlet and $I
\! = \! 0$, spin-triplet channels.  The positive fit value of
$V_0(1S)$ in Eq.~(\ref{eq:LastPaper}) suggests a similar pattern of
isospin coupling for the $S$-wave $\de$-$\bde$ states.  In considering
$P$-wave and higher states, we also introduce the isospin-dependent
tensor operator,
\begin{equation} \label{eq:TensorHam}
\Delta H_T = V_T \, {\bm \tau}_q \! \cdot \! {\bm \tau}_{\bar q} \;
S_{12}^{(\qq)} \, ,
\end{equation}
where the tensor operator $S_{12}$ is defined by
\begin{equation}
\label{eq:Tensor}
S_{12} \equiv 3 \, {\bm \sigma}_1 \! \cdot {\bm r} \, {\bm \sigma}_2
\! \cdot {\bm r} / r^2 - {\bm \sigma}_1 \! \cdot {\bm \sigma}_2 \, .
\end{equation}
For the purpose of this work, we consider only the tensor operator
$S_{12}^{(\qq)}$, in which the spin operators in Eq.~(\ref{eq:Tensor})
refer only to the light quarks, not the full diquarks.  We argue, just
as after Eq.~(\ref{eq:IsospinDiquark}), that the dominant
isospin-dependent tensor operator should be sensitive only to
light-quark spins; nevertheless, for completeness we tabulate in
Appendix~\ref{sec:Tensor} the $P$-wave multiplet matrix elements for
the alternative tensor operator
\begin{equation} \label{eq:TensorHam2}
\Delta H^\prime_T = V^\prime_T \, {\bm \tau}_q \! \cdot \!
{\bm \tau}_{\bar q} \; S_{12}^{(\de \bde)} \, .
\end{equation}

The matrix elements of $S_{12}$ are well known in the literature ({\it
e.g.}, Ref.~\cite{Shalit:1963nuclear}).  Defining for the generic case
${\bf J} \! \equiv \! {\bf L} \! + {\bf S}$, the distinct values are:
\begin{widetext}
\begin{eqnarray}
S_{12} \left| L = J, S=0, J, J_z \right> &=& 0 \, , \nonumber \\
\left< L' = J, S=1,J, J_z \right| S_{12} \left| L = J, S=1, J, J_z
\right> &=& 2 \, , \nonumber \\
\left< L' = J \! - \! 1, S=1,J, J_z \right| S_{12} \left| L = J \! -
\! 1, S=1, J, J_z
\right> &=&
-\frac{2\left(J -1\right)}{2J+1} \, , \nonumber \\
\left< L' = J \! + \! 1, S=1,J, J_z \right| S_{12} \left| L = J \! +
\! 1, S=1, J, J_z
\right> &=&
-\frac{2\left(J +2\right)}{2J +1} \, , \nonumber \\
\left< L' = J \! + \! 1, S=1,J, J_z \right| S_{12} \left| L = J \! -
\! 1, S=1, J, J_z
\right> &=&
\frac{6\sqrt{J\left(J+1\right)}}{2J +1} \, .
\label{eq:S12simple}
\end{eqnarray}
\end{widetext}

For the tensor operator $S_{12}^{(\qq)}$ of Eq.~(\ref{eq:TensorHam}),
neither the basis in terms of $s_\qq, s_\QQ$ spins nor $s_\de, s_\bde$
spins is convenient for the computation of matrix elements.  Instead,
the recoupling of the orbital angular momentum directly to the
light-quark spin,
\begin{equation} \label{eq:JqqDef}
{\bf J}_\qq \equiv {\bf L} + {\bf s}_\qq \, ,
\end{equation}
is necessary.\footnote{This procedure implicitly assumes no orbital
angular momentum within either $\de$ or $\bde$, so that $L_\qq \! = \!
L$.}  The amplitudes for this recoupling are given by $6j$ symbols:
\begin{eqnarray}
\lefteqn{\mathcal{M}_{J_\qq} \equiv
\left< (L,s_\qq),J_\qq,s_\QQ, J | L,(s_{\qq},s_{\QQ}), S, J \right>
} & & \nonumber\\
= & & (-1)^{L+s_\qq+s_\QQ+J}\sqrt{[J_\qq][S]}
\left\{ \begin{array}{ccc}
L & s_\qq & J_\qq\\
s_\QQ & J & S 
\end{array} \right\} , \label{eq:JqqCoef}
\end{eqnarray}
and are tabulated for the states of the $P$-wave multiplet in
Table~\ref{tab:PwaveStates}.

In summary, the full Hamiltonian adopted for the 28 isomultiplets of
the $1P$ multiplet is the sum of the 5 operators in
Eqs.~(\ref{eq:Ham}), (\ref{eq:SpinOrbit}), and (\ref{eq:TensorHam}):
\begin{eqnarray}
H & = & M_0 + 2 \kappa_{qQ} ({\bf s}_q \! \cdot \! {\bf s}_Q +
{\bf s}_{\bar q} \! \cdot \! {\bf s}_{\bar Q}) + V_{LS} \,
{\bf L} \cdot {\bf S} \nonumber \\ & & + V_0 \, {\bm \tau}_q
\! \cdot \! {\bm \tau}_{\bar q} \; {\bm \sigma}_q \! \cdot \!
{\bm \sigma}_{\bar q} + V_T \, {\bm \tau}_q \! \cdot
\! {\bm \tau}_{\bar q} \; S_{12}^{(\qq)} \, .
\label{eq:FullHam}
\end{eqnarray}
We now present the mass expressions for the $1P$ states, listed in the
same order as in Table~\ref{tab:PwaveStates}, for both $I \! = \! 0$
and $I \! = \! 1$.  Matrices indicate sets of states degenerate in
$J^{PC}$; for example, the mixing matrix element between the $I \! =
\! 0$, $1^{--}$ states $\tilde{X}_0^{(1)}$ and $X_1^{(1)}$ is $-9V_T
\sqrt{3/5}$.  Equations~(\ref{M_1_m_m_I_0})--(\ref{M_2_m_p_I_1}),
written in the basis of good total heavy-quark and light-quark spin
eigenvalues, represent the central theoretical results of this work;
in the next section we confront them with existing data.
\begin{widetext}
\begin{eqnarray}
M_{1^{--}}^{I=0}&=& M_0 \begin{pmatrix} 1 & 0 & 0 & 0\\ 0 & 1 & 0 &
0\\ 0 & 0 & 1 & 0\\ 0 & 0 & 0 & 1
\end{pmatrix}
+\kappa_{qQ}
\begin{pmatrix}
0 & -\sqrt{3} & 0 & 0 \\
-\sqrt{3} & -2 & 0 & 0 \\
0 & 0 & -1 & 0 \\
0& 0 & 0 & 1 \\
\end{pmatrix}
-V_{LS}
\begin{pmatrix}
0 & 0 & 0 & 0\\
0 & 0 & 0 & 0\\
0 & 0 & 1 & 0\\
0 & 0 & 0 & 3
\end{pmatrix}
-3 V_0
\begin{pmatrix}
-3 & 0 & 0 & 0\\
0 & 1 & 0 & 0\\
0 & 0 & 1 & 0\\
0 & 0 & 0 & 1
\end{pmatrix} \nonumber\\
&&-3V_T
\begin{pmatrix}
0 & 0 & 0 & 0 \\
0 & 0 & 0 & -\frac{4}{\sqrt{5}} \\
0 & 0 & -1 & 3\sqrt{\frac{3}{5}} \\
0 & -\frac{4}{\sqrt{5}} &  3\sqrt{\frac{3}{5}} & -\frac{7}{5} \\
\end{pmatrix} \, , \label{M_1_m_m_I_0}
\end{eqnarray}
\begin{eqnarray}
M_{1^{--}}^{I=1} &=&M_0 \begin{pmatrix}
1 & 0 & 0 & 0\\
0 & 1 & 0 & 0\\
0 & 0 & 1 & 0\\
0 & 0 & 0 & 1
\end{pmatrix}
+\kappa_{qQ}
\begin{pmatrix}
0 & -\sqrt{3} & 0 & 0 \\
-\sqrt{3} & -2 & 0 & 0 \\
0 & 0 & -1 & 0 \\
0& 0 & 0 & 1 \\
\end{pmatrix}
-V_{LS}
\begin{pmatrix}
0 & 0 & 0 & 0\\
0 & 0 & 0 & 0\\
0 & 0 & 1 & 0\\
0 & 0 & 0 & 3
\end{pmatrix}
+ V_0
\begin{pmatrix}
-3 & 0 & 0 & 0\\
0 & 1 & 0 & 0\\
0 & 0 & 1 & 0\\
0 & 0 & 0 & 1
\end{pmatrix} \nonumber\\
&& +V_T
\begin{pmatrix}
0 & 0 & 0 & 0 \\
0 & 0 & 0 & -\frac{4}{\sqrt{5}} \\
0 & 0 & -1 & 3\sqrt{\frac{3}{5}} \\
0 & -\frac{4}{\sqrt{5}} &  3\sqrt{\frac{3}{5}} & -\frac{7}{5} \\
\end{pmatrix} \, , \label{M_1_m_m_I_1}
\end{eqnarray}
\begin{eqnarray}
M_{0^{--}}^{I=0} &=& M_0 -\kappa_{qQ}-2V_{LS}-3V_0-6V_T \, ,
\label{M_0_m_m_I_0}\\
M_{0^{--}}^{I=1} &=& M_0 -\kappa_{qQ}-2V_{LS}+V_0+2V_T \, ,
\label{M_0_m_m_I_1}\\
M_{2^{--}}^{I=0} &=& M_0\begin{pmatrix}
1 & 0\\
0 & 1
\end{pmatrix}
+\kappa_{qQ}
\begin{pmatrix}
-1 & 0\\
0 & 1
\end{pmatrix}
+V_{LS}
\begin{pmatrix}
1 & 0 \\
0 & -1
\end{pmatrix}
-3V_0
\begin{pmatrix}
1 & 0 \\
0 & 1
\end{pmatrix}
-\frac{3}{5} V_T
\begin{pmatrix}
1 & -3\sqrt{3} \\
-3\sqrt{3} & 7
\end{pmatrix} \, , \label{M_2_m_m_I_0}\\
M_{2^{--}}^{I=1} &=& M_0\begin{pmatrix}
1 & 0\\
0 & 1
\end{pmatrix}
+\kappa_{qQ}
\begin{pmatrix}
-1 & 0\\
0 & 1
\end{pmatrix}
+V_{LS}
\begin{pmatrix}
1 & 0 \\
0 & -1
\end{pmatrix}
+V_0
\begin{pmatrix}
1 & 0 \\
0 & 1
\end{pmatrix}
+\frac{1}{5} V_T
\begin{pmatrix}
1 & -3\sqrt{3} \\
-3\sqrt{3} & 7
\end{pmatrix} \, , \label{M_2_m_m_I_1}\\
M_{3^{--}}^{I=0} &=& M_0 +\kappa_{qQ}+2V_{LS}-3V_0+\frac{6}{5}V_T
\, , \label{M_3_m_m_I_0}\\
M_{3^{--}}^{I=1} &=& M_0 +\kappa_{qQ}+2V_{LS}+V_0-\frac{2}{5}V_T
\, , \label{M_3_m_m_I_1}\\
M_{0^{-+}}^{I=0} &=& M_0\begin{pmatrix}
1 & 0 \\
0 & 1 
\end{pmatrix}
+\kappa_{qQ}
\begin{pmatrix}
0 & 1 \\
1 & 0 
\end{pmatrix}
-2V_{LS}
\begin{pmatrix}
1 & 0 \\
0 & 1 \\
\end{pmatrix}
-3V_0
\begin{pmatrix}
-3 & 0 \\
0 & 1 
\end{pmatrix}
+12V_T
\begin{pmatrix}
1 & 0 \\
0 & 0 \\
\end{pmatrix} \, , \label{M_0_m_p_I_0}\\
M_{0^{-+}}^{I=1} &=& M_0\begin{pmatrix}
1 & 0 \\
0 & 1 
\end{pmatrix}
+\kappa_{qQ}
\begin{pmatrix}
0 & 1 \\
1 & 0 
\end{pmatrix}
-2V_{LS}
\begin{pmatrix}
1 & 0 \\
0 & 1 \\
\end{pmatrix}
+V_0
\begin{pmatrix}
-3 & 0 \\
0 & 1 
\end{pmatrix}
-4V_T
\begin{pmatrix}
1 & 0 \\
0 & 0 \\
\end{pmatrix} \, , \label{M_0_m_p_I_1}\\
M_{1^{-+}}^{I=0} &=& M_0\begin{pmatrix}
1 & 0 \\
0 & 1 
\end{pmatrix}
+\kappa_{qQ}
\begin{pmatrix}
0 & 1 \\
1 & 0 
\end{pmatrix}
-V_{LS}
\begin{pmatrix}
1 & 0 \\
0 & 1 
\end{pmatrix}
-3V_0
\begin{pmatrix}
-3 & 0 \\
0 & 1 
\end{pmatrix}
-6V_T
\begin{pmatrix}
1 & 0 \\
0 & 0 \\
\end{pmatrix} \, , \label{M_1_m_p_I_0}\\
M_{1^{-+}}^{I=1} &=& M_0\begin{pmatrix}
1 & 0 \\
0 & 1 
\end{pmatrix}
+\kappa_{qQ}
\begin{pmatrix}
0 & 1 \\
1 & 0 
\end{pmatrix}
-V_{LS}
\begin{pmatrix}
1 & 0 \\
0 & 1 
\end{pmatrix}
+V_0
\begin{pmatrix}
-3 & 0 \\
0 & 1 
\end{pmatrix}
+2V_T
\begin{pmatrix}
1 & 0 \\
0 & 0 \\
\end{pmatrix} \, , \label{M_1_m_p_I_1}\\
M_{2^{-+}}^{I=0} &=& M_0\begin{pmatrix}
1 & 0\\
0 & 1
\end{pmatrix}
+\kappa_{qQ}
\begin{pmatrix}
0 & 1\\
1 & 0
\end{pmatrix}
+V_{LS}
\begin{pmatrix}
1 & 0 \\
0 & 1
\end{pmatrix}
-3V_0
\begin{pmatrix}
-3 & 0 \\
0 & 1
\end{pmatrix}
+\frac{6}{5} V_T
\begin{pmatrix}
1 & 0 \\
0 & 0 
\end{pmatrix} \, , \label{M_2_m_p_I_0}\\
M_{2^{-+}}^{I=1} &=& M_0\begin{pmatrix}
1 & 0\\
0 & 1
\end{pmatrix}
+\kappa_{qQ}
\begin{pmatrix}
0 & 1\\
1 & 0
\end{pmatrix}
+V_{LS}
\begin{pmatrix}
1 & 0 \\
0 & 1
\end{pmatrix}
+V_0
\begin{pmatrix}
-3 & 0 \\
0 & 1
\end{pmatrix}
-\frac{2}{5} V_T
\begin{pmatrix}
1 & 0 \\
0 & 0 
\end{pmatrix} \, . \label{M_2_m_p_I_1}
\end{eqnarray}
\end{widetext}

\section{Comparison to Data}
\label{sec:Analysis}

The experimental evidence at the present time for a complete
multiplet of $P \! = \! -$ hidden-charm $4$-quark exotic states is
fragmentary, as should be apparent from the overview of
Sec.~\ref{sec:ExptReview}.  Only between 3--6 such states out of 28
in $\Sigma^+_g(1P)$ have been clearly identified, most of which have
$J^{PC} \!  = \! 1^{--}$.  Moreover, according to
Table~\ref{tab:PwaveStates} and
Eqs.~(\ref{M_1_m_m_I_0})--(\ref{M_1_m_m_I_1}), this sector possesses
a fourfold multiplicity for both $I \! = \! 0$ and $I \! = \! 1$, and
therefore the exact $1^{--}$ mass eigenvalues depend nontrivially
upon the values of the 5 model parameters $M_0, \kappa_{qQ}, V_{LS},
V_0, V_T$ of the Hamiltonian in Eq.~(\ref{eq:FullHam}).  In addition,
if this model lacks an operator---even one with a fairly small
coefficient--- that produces a pattern of mass matrix elements
distinct from the pattern already present in
Eqs.~(\ref{M_1_m_m_I_0})--(\ref{M_1_m_m_I_1}), then the effect on the
spectrum of mass eigenvalues could be quite pronounced.

One could avoid such problems by constraining not the individual
eigenvalues directly, but rather the coefficients of the
characteristic polynomial of the $1^{--}$ mass matrix $M_{1^{--}}$,
such as its trace.  However, obtaining constraints in this way
requires one to be confident of which 4 states actually belong to the
$1P$ multiplet (more on this below).  In addition, the factors of
${\rm Tr} \, M^N_{1^{--}}$ that comprise the characteristic
polynomial are numerically dominated by $M_0(1P)^N$ and are less
sensitive to mass splittings.  Furthermore, each factor of ${\rm Tr}
\, M^N_{1^{--}}$ treats all eigenvalues symmetrically, while in
practice one possesses indications from the data that a particular
eigenvalue must be matched to a particular eigenvector representing a
state with a particular pattern of decays as indicated by the quantum
numbers of particles in the final state (again, more on this below).

As a result, the fits presented here are exploratory in nature, and
depend crucially upon provisional assignments of observed states to
particular roles in the model.  Conversely, with so little data
completely settled, this model provides definite predictions for the
full spectrum of states, given any specific hypothesis for the
identity of states thus far observed.

Even if a $1^{--}$ state is experimentally confirmed, its membership
in the $1P$ multiplet can be in doubt.  Specifically, $Y(4660)$ is
well separated in mass from the other $1^{--}$ states, and has been
suggested as a $2P$ state~\cite{Giron:2019bcs} or a $1F$
state~\cite{Maiani:2014aja}.  Using the methods of
Ref.~\cite{Giron:2019bcs}, which produce the values of $M_0(1S)$ in
Eq.~(\ref{eq:LastPaper}) and $M_0(1P)$ in Eq.~(\ref{eq:Mass1P}), we
find
\begin{eqnarray}
M_0(2P) & = & 4819 \mbox{-} 4824 \ {\rm MeV} \, , \nonumber \\
M_0(1F) & = & 4886 \mbox{-} 4891 \ {\rm MeV} \, .
\end{eqnarray}
Since the $2P$-$1P$ splitting is then approximately 460~MeV, and the
lightest $1P$ candidate is $Y(4230)$ at $4218^{+5}_{-4}$~MeV, one may
crudely estimate the lightest $2P$ candidate to lie around
$4680$~MeV, which is not excessively higher than the measured value
of $4643 \! \pm \! 9$~MeV\@.  The $1F$ assignment, on the contrary,
appears to be out of reach for $Y(4660)$.  Even though $2P$ quantum
numbers certainly present a likely possibility, one may alternately
consider (as is done below) a fit in which $Y(4660)$ is taken to be
the highest of the 4 $1P$, $1^{--}$ states.

If $Y(4660)$ is assumed to be an $I \! = \! 0$, $1P$ state, then one
of $Y(4230)$, $Y(4260)$, $Y(4360)$, $Y(4390)$ must be superfluous.  As
discussed in Sec.~\ref{sec:ExptReview}, BESIII sees no strong evidence
in the $J/\psi \, \pi \pi$ channel near 4260~MeV of $Y(4260)$ (which
interestingly, was historically the first of the $Y$ states
observed~\cite{Aubert:2005rm}); rather, their measurement ascribed by
the PDG to $Y(4260)$ agrees quite well with the value given for
$Y(4230)$, and the latest BESIII value~\cite{Ablikim:2019apl} agrees
even better.  One may then either suppose that all of the other
$Y(4260)$ data, obtained via the process $e^+ e^- \! \to \! \gamma Y$,
represent a $4$-quark state that is for some reason inaccessible to
direct production via $e^+ e^- \! \to \! Y$, or it is not an
independent $4$-quark state at all, being instead an obsolete
experimental artifact properly subsumed into $Y(4230)$, or possibly
even a signal of the low-lying $1^{--}$ charmonium hybrid expected
near that mass~\cite{Liu:2012ze}.

Recall also from Sec.~\ref{sec:ExptReview} the interesting fact that
the two BESIII measurements of $m_{Y(4360)}$ are very widely
separated, with the higher mass being completely consistent with that
of $Y(4390)$ (although its width differs by about $2\sigma$), and the
lower one near 4320~MeV suggesting a new state, possibly subsuming
what had been the high-mass tail of $Y(4260)$ events in other
experiments.  One may then consider a BESIII-only set of $1^{--}$
states, given by $Y(4230)$, ``$Y(4320)$'', and $Y(4390)$, and then
either predict a fourth $1^{--}$ mass, or include $Y(4660)$ as the
fourth $1^{--}$ state in $1P$.

The only other known $P \! = \! -$ candidate is the unconfirmed
$Z_c(4240)$, whose $I \! = \! 1$, $J^{PC} \! = 0^{--}$ quantum numbers
are unique in the $1P$ multiplet.  Since its mass uncertainty (see
Table~\ref{tab:Expt}) is quite large, we choose to predict
$m_{Z_c(4240)}$ from the fits rather than use it as an input.

Apart from the spectrum of mass eigenvalues, the most incisive
observed feature of $1^{--}$ states is that at least 2 of them,
$Y(4230)$ and $Y(4390)$, have been seen to decay to the $s_{\QQ} \! =
\! 0$ charmonium state $h_c$.  In addition, $Y(4230)$ is also seen to
decay to the $s_{\QQ} \! = \! 1$ states $\chi_{c0}$ and $\psi(2S)$,
indicating a significant mixing of $s_{\QQ} \! = \! 0$ and $s_{\QQ}
\! = \! 1$ components within $Y(4230)$ (assuming that heavy-quark
spin is conserved in the decays).  The $Y(4390)$, on the other hand,
has thus far only been seen to decay to $h_c$, although if the higher
BESIII mass measurement $4384 \! \pm \! 4$~MeV for $e^+ e^- \! \to \!
\psi(2S) \pi^+ \pi^-$\cite{Ablikim:2017oaf} ascribed to $Y(4360)$
actually belongs to $Y(4390)$, then this state also comprises a
mixture of both $s_{\QQ}$ components.  But in any case, at least 2 of
the $1^{--}$ states have a significant coupling to $s_{\QQ} \! = \!
0$, while Table~\ref{tab:PwaveStates} indicates only 1 such state
($\tilde X^{(1)}_0$) for each of $I \! = \! 0$ and $I \! = \! 1$.  The
$I \! = \! 0$ $Y(4230)$ is clearly one of them, while $Y(4390)$ may be
considered to carry either $I \! = \! 0$ or $I \! = \!  1$ (although
again, recent BESIII $J/\psi \, \eta$
data~\cite{ablikim2020observation} appear to eliminate the latter
possibility).  The probability $P_{s_{\QQ} = 0}$ of coupling to the
$s_{\QQ} \! = \! 0$ component of a mass eigenstate is given by the
square of the $\tilde X^{(1)}_0$ component of the normalized mass
eigenvector.

We perform $\chi^2$ fits and compute $\chi^2_{\rm min}$ for a given
set of inputs $M_0, \kappa_{qQ}, V_{LS}, V_0, V_T$ by including terms
for mass measurements and their uncertainties in the standard manner,
while terms for $P_{s_{\QQ} = 0}$ are included in the form
\begin{equation}
\Delta \chi^2 = \left( \frac{\ln P_{s_{\QQ} = 0} - \ln f}{\epsilon}
\right)^2 \ ,
\end{equation}
so that $P_{s_{\QQ} = 0}$ is fit to a chosen branching fraction $f$,
values $f \! \to \! 0$ thus being disfavored, and one unit of $\chi^2$
for this variable is approximately bracketed by the values $P_{s_{\QQ}
= 0} \! = \! f(1 \! \pm \! \epsilon)$.  The exact branching fractions
into $s_{\QQ} \! = \! 0$ and $s_{\QQ} \! = \! 1$ states are not yet
known, so these constraints upon the $\chi^2$ function should be
viewed as typical possible values rather than precise conditions
imposed by the data.  In particular, in light of the decay modes for
$Y(4230)$ and $Y(4390)$ listed in Table~\ref{tab:Expt}, some explicit
examples we explore include fixing a substantial $s_{\QQ = 0}$
component of $f \! = \! \frac 1 3$ for $Y(4230)$ and a large $s_{\QQ =
0}$ component of $f \!  = \! \frac 2 3$ for $Y(4390)$.

Having now motivated a diverse set of alternative ways to interpret
the data for the $1^{--}$ states, we now define a variety of specific
scenarios to test against the model Hamiltonian of
Eq.~(\ref{eq:FullHam}), using $\chi^2$ fits in the manner just
outlined.  Clearly, not every combination of possible interpretations
of the data as described above is represented by these 5 cases, but
they are useful in understanding how well various fits to
Eq.~(\ref{eq:FullHam}) succeed in representing the full extant body of
experimental results.  The results of these fits are presented in
Table~\ref{tab:Cases}.  Explicitly, the cases are:

\begin{enumerate}

\item $Y(4230)$, $Y(4260)$, $Y(4360)$, $Y(4390)$ masses are as given in
the PDG (Table~\ref{tab:Expt}).  No constraint is placed upon
$P_{s_{\QQ} = 0}^{Y(4230)}$ or $P_{s_{\QQ} =
0}^{Y(4390)}$. \label{JunkFit19}

\item $Y(4230)$, $Y(4260)$, $Y(4360)$, $Y(4390)$ masses are as given in
the PDG\@.  $P_{s_{\QQ} = 0}^{Y(4230)}$ is fit to $f \! = \! \frac 1
3$ with $\epsilon \! = \! 0.1$, and $P_{s_{\QQ} = 0}^{Y(4390)}$ is
unconstrained. \label{JunkFit7B}

\item $Y(4230)$, $Y(4360)$, and $Y(4390)$ masses are as given in
the PDG, while $m_{Y(4260)} \! = \! 4251 \! \pm \! 6$~MeV, which is
the weighted average of the 3 PDG values not including the low BESIII
value~\cite{Ablikim:2016qzw}.  $P_{s_{\QQ} = 0}^{Y(4230)}$ is fit to
$f \! = \! \frac 1 3$ with $\epsilon \! = \! 0.2$, and $P_{s_{\QQ} =
0}^{Y(4390)}$ is fit to $f \! = \! \frac 2 3$ with $\epsilon \! = \!
0.05$. \label{JunkFit9B}

\item $Y(4360)$, $Y(4390)$, and $Y(4660)$ masses are as
given in the PDG, but $Y(4260)$ is assumed not to exist, and
$m_{Y(4230)} \!  = \! 4220.1 \! \pm 2.9$~MeV is the weighted average
of the PDG values combined with the newer BESIII
measurements~\cite{Ablikim:2018vxx,Ablikim:2019apl}.  $P_{s_{\QQ} =
0}$ values are as given in Case~\ref{JunkFit9B}. \label{JunkFit21B}

\item $m_{Y(4230)}$ is as given in Case~\ref{JunkFit21B};
$m_{Y(4260)}$ is as given in Case~\ref{JunkFit9B};
$m_{\text{``$Y(4320)$''}} \! = \! 4320 \! \pm \! 13$~MeV is the lower
BESIII $Y(4360)$ mass measurement from~\cite{Ablikim:2016qzw};
$m_{Y(4390)} \! = \! 4386 \! \pm \! 4$~MeV is the weighted average of
the PDG value and the upper BESIII mass measurement
from~\cite{Ablikim:2017oaf}.   $P_{s_{\QQ} = 0}$ values are as given
in Case~\ref{JunkFit9B}.
\label{JunkFit18B}

\end{enumerate}

\begin{table*}[ht]
\centering
\caption{Results of fits to the Hamiltonian parameters $M_0,
\kappa_{qQ}, V_{LS}, V_0, V_T$ in Eq.~(\ref{eq:FullHam}) using the
$1^{--}$ state assigments and masses summarized in the text as
Cases~\ref{JunkFit19}--\ref{JunkFit18B}.  The mass of the $0^{--}$,
$I \! = \! 1$ state, to be compared with $m_{Z_c(4240)}$, is also
predicted.  All Hamiltonian parameters and mass predictions are given
in units of MeV\@.  Also presented are fractional amounts $P_{s_{\QQ}
= 0}$ of the heavy-quark spin $s_{\QQ} \! = \! 0$ state in the mass
eigenstates that correspond to $Y(4230)$ and $Y(4390)$.}
\setlength{\extrarowheight}{1.3ex}
\label{tab:Cases}
\begin{tabular}{c c c c r r r @{ \ } c c c c @{ \ } c c c}
\hline\hline
 & $\chi^2_{\rm min}/\rm{d.o.f.}$ & $M_0$ & $\kappa_{qQ}$ & $V_{LS}$ & $V_0 \ $ & $V_T$ & \multicolumn{4}{c}{$M_{1^{--}}^{I=0}$} & $M_{0^{--}}^{I=1}$ & $P_{s_{\QQ} = 0}^{Y(4230)}$ & $P_{s_{\QQ} = 0}^{Y(4390)}$\\
\hline
Case 1 & $0.457/4$ & $4356$ & $12.7$ & $61.6$ & $-14.7$ & $7.0$ & $4218.2$ & $4230.4$ & $4360.0$ & $4393.9$ & $4219.4$ & $0.771$ & $0.012$\\
Case 2 & $2.47/5$ & $4355$ & $19.9$ & $59.8$ & $-12.9$ & $8.3$ & $4217.3$ & $4237.9$ & $4351.4$ & $4393.3$ & $4219.2$ & $0.331$ & $0.023$\\
Case 3 & $15.2/6$ & $4357$ & $44.6$ & $42.9$ & $-1.8$ & $5.5$ & $4213.6$ & $4262.9$ & $4335.5$ & $4394.8$ & $4235.0$ & $0.231$ & $0.639$\\
Case 4 & $5.33/6$ & $4357$ & $43.7$ & $-53.7$ & $-0.2$ & $10.4$ & $4219.3$ &$4374.5$ & $4399.2$ & $4637.6$ & $4441.4$ & $0.232$ & $0.647$\\
Case 5 & $3.76/6$ & $4356$ & $43.2$ & $49.0$ & $-2.7$ & $3.8$ & $4219.3$ &$4257.1$ & $4306.6$ & $4385.7$ & $4219.7$ & $0.264$ & $0.651$\\
\hline\hline
\end{tabular}
\end{table*}

The robustness of the $\chi^2_{\rm min}$ values are checked by
selecting several initial sets of parameters and confirming that the
same minimum is reached for each such choice.  Neither uncertainties
on the output parameters nor a covariance matrix are presented for the
results of Table~\ref{tab:Cases}, partly since the $1^{--}$ mass
inputs are eigenvalues of a $4 \! \times \! 4$~matrix, which satisfy a
highly nonlinear characteristic equation.  More broadly, however,
presenting such detailed fits to data that remain strongly in flux
would suggest more confidence in the exact fit values of the output
parameters than is warranted at this time.

The first feature of note in Table~\ref{tab:Cases} is that the values
of $\chi^2_{\rm min}$ are entirely satisfactory for
Cases~\ref{JunkFit19}, \ref{JunkFit7B},
\ref{JunkFit21B}, and \ref{JunkFit18B}, and large for
Case~\ref{JunkFit9B}.  These results do not simply mean that some
assignments of states are intrinsically better than others; rather,
they indicate which pieces of data and which parameters are
instrumental in driving quality of the fit, as discussed below.

The fit values of $M_0(1P)$ are remarkably robust, even in
Case~\ref{JunkFit21B}, which treats $Y(4660)$ as the highest $1P$
state.  Note that the $M_0(1P)$ value is {\em not\/} taken from
Eq.~(\ref{eq:Mass1P}), but rather emerges via the $\chi^2$
optimization.
The price of treating $Y(4660)$ as a $1P$ rather than a $2P$ state is
the prediction that the sole $I \!  = \! 1$, $0^{--}$ state in the
$1P$ multiplet is much heavier than the observed $Z_c(4240)$, while
all of the other fits provide satisfactory values for $m_{Z_c(4240)}$.
The value of $V_{LS}$ is also rather stable in all cases that do not
include $Y(4660)$; $V_{LS}$ appears to be the primary parameter
responsible for the largest mass splittings within the $1P$ multiplet.

If $Y(4660)$ is instead required to be a $2P$ state, then one may
begin with the minimal assumption in Case~\ref{JunkFit19} that the 4
$1P$ $Y$ states are just the lighter $1^{--}$ ones listed in the PDG,
and the mixings $P_{s_{\QQ} = 0}$ are allowed to vary freely.  One
finds, unsurprisingly, a perfect fit to the masses, and that almost
all of the $s_{\QQ} \! = \! 0$ strength resides with $Y(4230)$, but
almost none with $Y(4390)$, contrary to observation.  The value of
$\kappa_{qQ}$ obtained actually turns out to be smaller than the value
obtained in Eq.~(\ref{eq:LastPaper}) for the $1S$ multiplet.

However, when one also fits to $P_{s_{\QQ} = 0}^{Y(4230)} \! = \!
\frac 1 3$ [Case~\ref{JunkFit7B}], the $\chi^2_{\rm min}/{\rm
d.o.f.}$ rises somewhat---still providing a good fit---but most
significantly, $\kappa_{qQ}$ becomes almost equal to the value in
Eq.~(\ref{eq:LastPaper}).  If one then also attempts to fit to
$P_{s_{\QQ} = 0}^{Y(4230)} \! = \! \frac 2 3$ as well (a case not
presented in Table~\ref{tab:Cases}), then the value of $\chi^2_{\rm
min}/{\rm d.o.f.}$ rises dramatically, its increase driven by the
tendency of the fit to prefer a larger $m_{Y(4260)}$ and a smaller
$m_{Y(4360)}$ than the PDG values.

Case~\ref{JunkFit9B} represents a halfway point designed to relieve
this tension; $m_{Y(4260)}$ is taken to be the average obtained from
the larger, non-BESIII PDG measurements, with the smaller BESIII
measurement subsumed into $m_{Y(4230)}$.  One finds the resulting
value of $\kappa_{qQ}$ to be much larger, a feature apparently
necessary to provide substantial $P_{s_{\QQ} = 0}$ values for both
$Y(4230)$ and $Y(4390)$, and the larger $\chi^2_{\rm min}/{\rm
d.o.f.}$ value is driven primarily by the preference of the fit for a
smaller $m_{Y(4360)}$.

We therefore modify Case~\ref{JunkFit9B} to consider in
Case~\ref{JunkFit18B} the interesting possibility discussed above
(inspired by BESIII measurements) that the $Y(4360)$ data actually
represent a combination of data that belongs to $Y(4390)$ and a
lighter state ``$Y(4320)$''.  One then obtains an excellent fit to all
observables.  The stable values of $M_0(1P)$, $V_{LS}$, and large
value of $\kappa_{qQ}$ have already been noted.  In addition, most of
the fits in Table~\ref{tab:Cases} prefer small, generally negative
values of $V_0$, quite different from the $1S$ value in
Eq.~(\ref{eq:LastPaper}), suggesting a rather different
isospin-dependent interaction in the $1P$ states.  Still, one must
note that the $V_0$ and $V_T$ couplings in the Hamiltonian of
Eq.~(\ref{eq:FullHam}) solely couple to static light quarks; the
presence of nonzero orbital angular momentum clearly changes the
physical interpretation one should apply to the corresponding
operators.

We do not insist that this particular choice of masses represents the
true spectrum of $I \! = \! 0$, $1^{--}$ $1P$ states that will
eventually be clarified by future experiments; in particular, our
simplistic estimates of $P_{s_{\QQ} = 0}$ values certainly skew the
fit results, and may not bear up under future scrutiny.  Nevertheless,
our purpose here is to show the versatility of the Hamiltonian of
Eq.~(\ref{eq:FullHam}) in accommodating data sets not unlike those
already obtained, with the key restriction of the dynamical diquark
model being that exactly 4 $I \! = \! 0$, $1^{--}$ states and 1 $I \!
= \! 1$, $0^{--}$ state occur in the $1P$ multiplet.

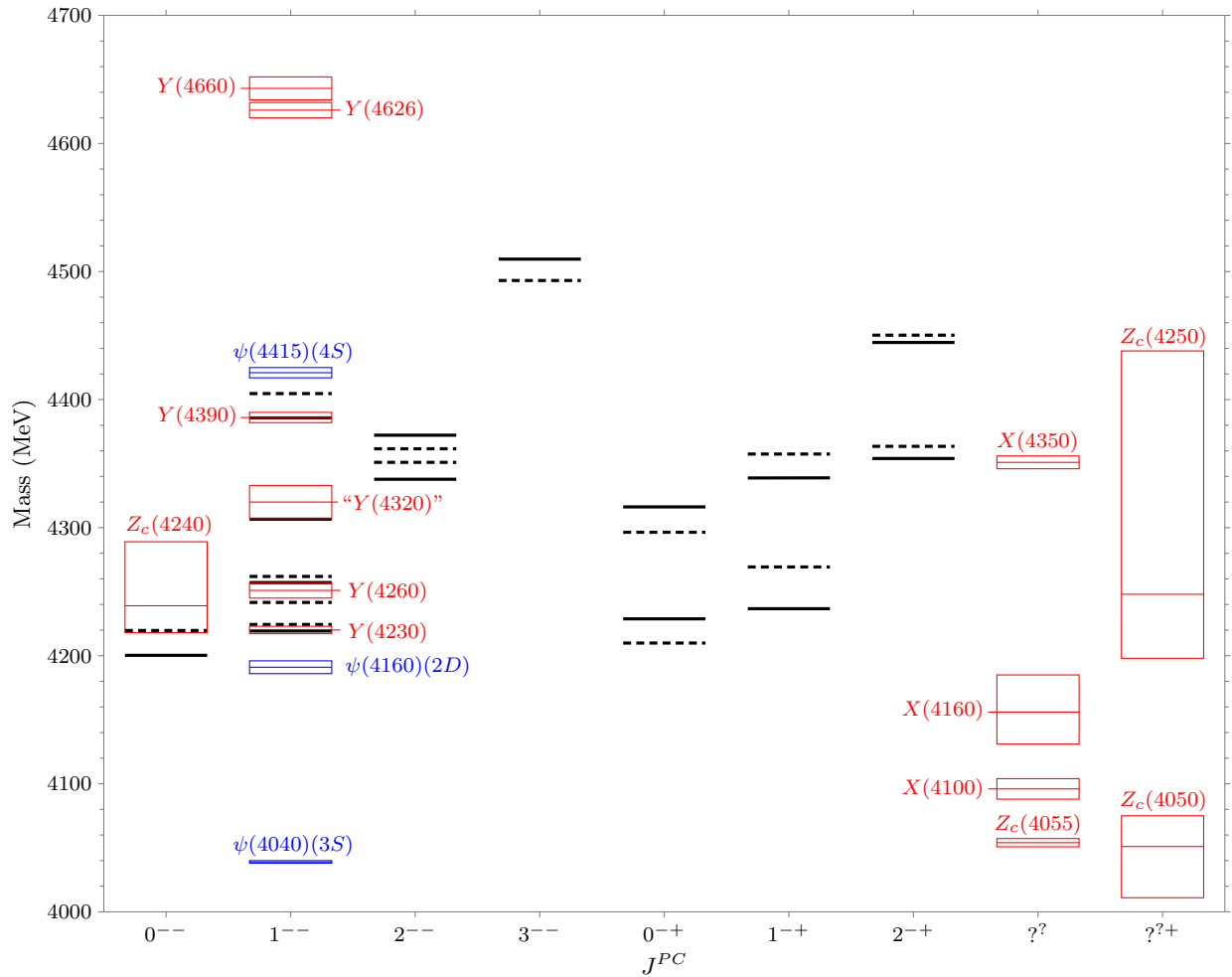
\begin{figure*}[t]
\begin{center}
\begin{tikzpicture}
\def\ncol{9} \def\mmin{4000} \def\mmax{4700} \def\mstep{100}
\def\w{0.33} \def\h{0.5}
\def\mD0{1864.83} \def\mDp{1869.65}
\def\mDstar0{2006.85} \def\mDstarp{2010.26}
\def\mDs{1968.34} \def\mDstars{2112.2} \def\mLambdac{2286.46}
\datavisualization[scientific axes={width=15.0cm, height=12.0cm},
    x axis={attribute=jpc, label={$J^{PC}$}, min value=\h,
    max value=(\ncol+\h), ticks={step=1,major at={
    1 as $0^{--}$, 2 as $1^{--}$, 3 as $2^{--}$, 4 as $3^{--}$,
    5 as $0^{-+}$, 6 as $1^{-+}$, 7 as $2^{-+}$, 8 as $?^{?}$,
    9 as $?^{?+}$}}},
    y axis={attribute=m, label={Mass (MeV)}, min value=\mmin,
    max value=\mmax, ticks={step=\mstep, minor steps between steps=4,
    style={/pgf/number format/set thousands separator=}}}]
    info {
      \draw [blue]
      (visualization cs: jpc={(2-\w)}, m={4039})  --
      (visualization cs: jpc={(2+\w)}, m={4039});
      \draw [blue]
      (visualization cs: jpc={(2-\w)}, m={(4039-1)})
      rectangle
      (visualization cs: jpc={(2+\w)}, m={(4039+1)})
      node [above=1.5ex, left=-3.0ex, font=\footnotesize] {$\psi(4040) (3S)$};
      \draw [red]
      (visualization cs: jpc={(9-\w)}, m={4051})  --
      (visualization cs: jpc={(9+\w)}, m={4051});
      \draw [red]
      (visualization cs: jpc={(9-\w)}, m={(4051+24)})
      rectangle
      (visualization cs: jpc={(9+\w)}, m={(4051-40)})
      node [above=9.5ex, left=-1.2ex, font=\footnotesize] {$Z_c (4050)$};
      \draw [red]
      (visualization cs: jpc={(8-\w)}, m={4054})  --
      (visualization cs: jpc={(8+\w)}, m={4054});
      \draw [red]
      (visualization cs: jpc={(8-\w)}, m={(4054+3.2)})
      rectangle
      (visualization cs: jpc={(8+\w)}, m={(4054-3.2)})
      node [above=2.2ex, left=-1.1ex, font=\footnotesize] {$Z_c (4055)$};
      \draw [red]
      (visualization cs: jpc={(8-0.07-\w)}, m={4096})  --
      (visualization cs: jpc={(8+\w)}, m={4096});
      \draw [red]
      (visualization cs: jpc={(8-\w)}, m={(4096+8)})
      rectangle
      (visualization cs: jpc={(8+\w)}, m={(4096-8)})
      node [left=8.5ex, below=-1.0ex, left, font=\footnotesize] {$X(4100)$};
      \draw [red]
      (visualization cs: jpc={(8-0.07-\w)}, m={4156})  --
      (visualization cs: jpc={(8+\w)}, m={4156});
      \draw [red]
      (visualization cs: jpc={(8-\w)}, m={(4156-25)})
      rectangle
      (visualization cs: jpc={(8+\w)}, m={(4156+29)})
      node [left=8.5ex, below=3.5ex, left, font=\footnotesize] {$X(4160)$};
      \draw [blue]
      (visualization cs: jpc={(2-\w)}, m={4191})  --
      (visualization cs: jpc={(2+\w)}, m={4191});
      \draw [blue]
      (visualization cs: jpc={(2-\w)}, m={(4191-5)})
      rectangle
      (visualization cs: jpc={(2+\w)}, m={(4191+5)})
      node [right=7.5ex, below=-1.3ex, font=\footnotesize] {$\psi(4160)(2D)$};
      \draw [black, very thick]
      (visualization cs: jpc={(1-\w)}, m={4200.3})  --
      (visualization cs: jpc={(1+\w)}, m={4200.3});
      \draw [black, very thick, style=densely dashed]
      (visualization cs: jpc={(1-\w)}, m={4219.7})  --
      (visualization cs: jpc={(1+\w)}, m={4219.7});
      \draw [black, very thick]
      (visualization cs: jpc={(2-\w)}, m={4219.3})  --
      (visualization cs: jpc={(2+\w)}, m={4219.3});
      \draw [black, very thick, style=densely dashed]
      (visualization cs: jpc={(2-\w)}, m={4224.4})  --
      (visualization cs: jpc={(2+\w)}, m={4224.4});
      \draw [black, very thick]
      (visualization cs: jpc={(2-\w)}, m={4257.1})  --
      (visualization cs: jpc={(2+\w)}, m={4257.1});
      \draw [black, very thick, style=densely dashed]
      (visualization cs: jpc={(2-\w)}, m={4241.6})  --
      (visualization cs: jpc={(2+\w)}, m={4241.6});
      \draw [black, very thick]
      (visualization cs: jpc={(2-\w)}, m={4306.6})  --
      (visualization cs: jpc={(2+\w)}, m={4306.6});
      \draw [black, very thick, style=densely dashed]
      (visualization cs: jpc={(2-\w)}, m={4261.9})  --
      (visualization cs: jpc={(2+\w)}, m={4261.9});
      \draw [black, very thick]
      (visualization cs: jpc={(2-\w)}, m={4385.7})  --
      (visualization cs: jpc={(2+\w)}, m={4385.7});
      \draw [black, very thick, style=densely dashed]
      (visualization cs: jpc={(2-\w)}, m={4404.7})  --
      (visualization cs: jpc={(2+\w)}, m={4404.7});
      \draw [black, very thick]
      (visualization cs: jpc={(3-\w)}, m={4337.8})  --
      (visualization cs: jpc={(3+\w)}, m={4337.8});
      \draw [black, very thick, style=densely dashed]
      (visualization cs: jpc={(3-\w)}, m={4351.0})  --
      (visualization cs: jpc={(3+\w)}, m={4351.0});
      \draw [black, very thick]
      (visualization cs: jpc={(3-\w)}, m={4372.2})  --
      (visualization cs: jpc={(3+\w)}, m={4372.2});
      \draw [black, very thick, style=densely dashed]
      (visualization cs: jpc={(3-\w)}, m={4361.6})  --
      (visualization cs: jpc={(3+\w)}, m={4361.6});
      \draw [black, very thick]
      (visualization cs: jpc={(4-\w)}, m={4509.7})  --
      (visualization cs: jpc={(4+\w)}, m={4509.7});
      \draw [black, very thick, style=densely dashed]
      (visualization cs: jpc={(4-\w)}, m={4493.0})  --
      (visualization cs: jpc={(4+\w)}, m={4493.0});
      \draw [black, very thick]
      (visualization cs: jpc={(5-\w)}, m={4228.8})  --
      (visualization cs: jpc={(5+\w)}, m={4228.8});
      \draw [black, very thick, style=densely dashed]
      (visualization cs: jpc={(5-\w)}, m={4209.9})  --
      (visualization cs: jpc={(5+\w)}, m={4209.9});
      \draw [black, very thick]
      (visualization cs: jpc={(5-\w)}, m={4316.2})  --
      (visualization cs: jpc={(5+\w)}, m={4316.2});
      \draw [black, very thick, style=densely dashed]
      (visualization cs: jpc={(5-\w)}, m={4296.4})  --
      (visualization cs: jpc={(5+\w)}, m={4296.4});
      \draw [black, very thick]
      (visualization cs: jpc={(6-\w)}, m={4236.7})  --
      (visualization cs: jpc={(6+\w)}, m={4236.7});
      \draw [black, very thick, style=densely dashed]
      (visualization cs: jpc={(6-\w)}, m={4269.3})  --
      (visualization cs: jpc={(6+\w)}, m={4269.3});
      \draw [black, very thick]
      (visualization cs: jpc={(6-\w)}, m={4338.8})  --
      (visualization cs: jpc={(6+\w)}, m={4338.9});
      \draw [black, very thick, style=densely dashed]
      (visualization cs: jpc={(6-\w)}, m={4357.5})  --
      (visualization cs: jpc={(6+\w)}, m={4357.5});
      \draw [black, very thick]
      (visualization cs: jpc={(7-\w)}, m={4354.0})  --
      (visualization cs: jpc={(7+\w)}, m={4354.0});
      \draw [black, very thick, style=densely dashed]
      (visualization cs: jpc={(7-\w)}, m={4363.5})  --
      (visualization cs: jpc={(7+\w)}, m={4363.5});
      \draw [black, very thick]
      (visualization cs: jpc={(7-\w)}, m={4444.6})  --
      (visualization cs: jpc={(7+\w)}, m={4444.6});
      \draw [black, very thick, style=densely dashed]
      (visualization cs: jpc={(7-\w)}, m={4450.3})  --
      (visualization cs: jpc={(7+\w)}, m={4450.3});
      \draw [red]
      (visualization cs: jpc={(2+0.07+\w)}, m={4220.1})  --
      (visualization cs: jpc={(2+\w)}, m={4220.1});
      \draw [red]
      (visualization cs: jpc={(2-\w)}, m={(4220.1+2.9)})
      rectangle
      (visualization cs: jpc={(2+\w)}, m={(4220.1-2.9)})
      node [right=5.5ex, below=-2.0ex, font=\footnotesize] {$Y(4230)$};
      \draw [red]
      (visualization cs: jpc={(2-\w)}, m={4251})  --
      (visualization cs: jpc={(2+0.07+\w)}, m={4251});
      \draw [red]
      (visualization cs: jpc={(2-\w)}, m={(4251-6)})
      rectangle
      (visualization cs: jpc={(2+\w)}, m={(4251+6)})
      node [right=5.5ex, below=-1.0ex, font=\footnotesize] {$Y(4260)$};
      \draw [red]
      (visualization cs: jpc={(8-\w)}, m={4351})  --
      (visualization cs: jpc={(8+\w)}, m={4351});
      \draw [red]
      (visualization cs: jpc={(8-\w)}, m={(4351-5)})
      rectangle
      (visualization cs: jpc={(8+\w)}, m={(4351+5)})
      node [above=1.4ex, left=-0.8ex, font=\footnotesize] {$X(4350)$};
      \draw [red]
      (visualization cs: jpc={(2-\w)}, m={4320})  --
      (visualization cs: jpc={(2+0.05+\w)}, m={4320});
      \draw [red]
      (visualization cs: jpc={(2-\w)}, m={(4320-13)})
      rectangle
      (visualization cs: jpc={(2+\w)}, m={(4320+13)})
      node [right=6.0ex, below=-0.0ex, font=\footnotesize] {``$Y(4320)$''};
      \draw [red]
      (visualization cs: jpc={(9-\w)}, m={4248})  --
      (visualization cs: jpc={(9+\w)}, m={4248});
      \draw [red]
      (visualization cs: jpc={(9-\w)}, m={(4248+190)})
      rectangle
      (visualization cs: jpc={(9+\w)}, m={(4248-50)})
      node [above=31.5ex, left=-1.2ex, font=\footnotesize] {$Z_c (4250)$};
      \draw [red]
      (visualization cs: jpc={(2-0.07-\w)}, m={4386})  --
      (visualization cs: jpc={(2+\w)}, m={4386});
      \draw [red]
      (visualization cs: jpc={(2-\w)}, m={(4386-4)})
      rectangle
      (visualization cs: jpc={(2+\w)}, m={(4386+4)})
      node [left=13.2ex, below=-1.5ex, font=\footnotesize] {$Y(4390)$};
      \draw [blue]
      (visualization cs: jpc={(2-\w)}, m={4421})  --
      (visualization cs: jpc={(2+\w)}, m={4421});
      \draw [blue]
      (visualization cs: jpc={(2-\w)}, m={(4421-4)})
      rectangle
      (visualization cs: jpc={(2+\w)}, m={(4421+4)})
      node [above=1.5ex, left=-3.0ex, font=\footnotesize] {$\psi(4415)(4S)$};
      \draw [red]
      (visualization cs: jpc={(2-\w)}, m={4626})  --
      (visualization cs: jpc={(2+0.07+\w)}, m={4626});
      \draw [red]
      (visualization cs: jpc={(2-\w)}, m={(4626-6)})
      rectangle
      (visualization cs: jpc={(2+\w)}, m={(4626+6)})
      node [right=10.0ex, below=0.7ex, left, font=\footnotesize] {$Y(4626)$};
      \draw [red]
      (visualization cs: jpc={(1-\w)}, m={4239})  --
      (visualization cs: jpc={(1+\w)}, m={4239});
      \draw [red]
      (visualization cs: jpc={(1-\w)}, m={(4239+50)})
      rectangle
      (visualization cs: jpc={(1+\w)}, m={(4239-21)})
      node [above=10.5ex, left=-1.4ex, font=\footnotesize] {$Z_c (4240)$};
      \draw [red]
      (visualization cs: jpc={(2-0.07-\w)}, m={4643})  --
      (visualization cs: jpc={(2+\w)}, m={4643});
      \draw [red]
      (visualization cs: jpc={(2-\w)}, m={(4643-9)})
      rectangle
      (visualization cs: jpc={(2+\w)}, m={(4643+9)})
      node [left=8.5ex, below=1.0ex, left, font=\footnotesize] {$Y(4660)$};
    };
\end{tikzpicture}
\end{center}
\caption{Level diagram for the 28 negative-parity states of the
  $\Sigma^+_g(1P)$ multiplet, with masses as given in
  Table~\ref{tab:AllMasses} and predicted using the $Y$-state
  assignment of Case~\ref{JunkFit18B}.  Heavy solid (dashed) lines
  indicate $I \! = \! 0$ ($I \! = \! 1$) states, respectively.  Each
  observed state mass, including its central value and uncertainty, is
  presented as a rectangle, with exotic candidates in red and
  conventional charmonium states in blue.  Observed states labeled
  $Z_c$ are all $I \! = \!  1$, and all others are $I \! = \!  0$.
  The final two columns present observed states for which not all
  $J^{PC}$ values are known, although in addition, $Z_c(4055)$ is $C
  \! = \! -$ and $X(4350)$ is $C \! = \! +$.
\label{fig:LevelDiagram}}
\end{figure*}

Using the fit values obtained for the parameters $M_0, \kappa_{qQ},
V_{LS}, V_0, V_T$ from any of the cases listed in
Table~\ref{tab:Cases}, one may predict the masses for all 28
isomultiplets within the $1P$ multiplet.  For concreteness, we use the
inputs of Case~\ref{JunkFit18B}, and present these results in
Table~\ref{tab:AllMasses}, with a corresponding level diagram in
Fig.~\ref{fig:LevelDiagram}\@.  The $I \! = \! 0$, $1^{--}$ states
(called $Y$ above, and labeled as $\psi$ by the PDG, see
Table~\ref{tab:PwaveStates}) all fit well with set of the mass
observations specified in Case~\ref{JunkFit18B}; as noted above, this
part of the fit can be successfully realized by any of the given
cases.  In addition, the predicted mass of the $I \! = \! 1$, $0^{--}$
state nicely agrees with the measured mass of $Z_c(4240)$, which has
precisely these quantum numbers.

\begin{table}
\centering
\caption{Prediction of the 28 isomultiplet masses (in MeV) of the
$\Sigma^+_g(1P)$ multiplet, using the Hamiltonian of
Eq.~(\ref{eq:FullHam}) and the numerical values of the parameters
obtained from fit to the state assignment of Case~5 as given in
Table~\ref{tab:Cases}.  Boldface indicates fit outputs for states
whose measured masses are used as inputs for the fit.\vspace{1ex}}
\label{tab:AllMasses}
\setlength{\extrarowheight}{0.5ex}
\begin{tabular}{c c c @{ \ } c c}
\hline\hline
$J^{PC}$ & \multicolumn{2}{c}{$I=0$} & \multicolumn{2}{c}{$I=1$} \\
\hline
$1^{--}$ & $\bf{4219.3}$ & $\bf{4257.1}$ & $4224.4$ & $4241.6$ \\ 
                & $\bf{4306.6}$ & $\bf{4385.7}$ & $4261.9$ & $4404.7$ \\
$0^{--}$ & $4200.3$ & & ${4219.7}$\\
$2^{--}$ & $4337.8$ & $4372.2$ & $4351.0$ & $4361.6$ \\
$3^{--}$ & $4509.7$ & & $4493.0$\\
$0^{-+}$ & $4228.8$ & $4316.2$ & $4209.9$ & $4296.4$ \\
$1^{-+}$ & $4236.7$ & $4338.8$ & $4269.3$ & $4357.5$ \\
$2^{-+}$ & $4354.0$ & $4444.6$ & $4363.5$ & $4450.3$ \\
\hline\hline
\end{tabular}
\end{table}

The $I \! = \! 1$, $1^{--}$ states (PDG designation: $R_{c1}$) are
particularly interesting, as the subset with $I_3 \! = \! 0$ should
also be produced in $e^+ e^-$ collisions, and as noted above, might be
mistaken for $Y$ states. Remarkably, in Table~\ref{tab:AllMasses} and
Fig.~\ref{fig:LevelDiagram} all such $I \! = \! 1$ states lie fairly
close to ones with $I \! = \! 0$ (and not necessarily in a one-to-one
fashion).  The signal of the highest one, at 4405~MeV, might even be
obscured by the conventional charmonium state $\psi(4415)$.  Part of
the confusion about $1^{--}$ resonance data might thus ultimately find
its origin in overlapping signals from states with distinct isospin
eigenvalues, which would only be completely separated either through
observing nearly-degenerate charged partners for the $I \! = \! 1$
states, or by discerning final states consisting of conventional
charmonium plus a light hadron state of a known isospin eigenvalue.

Beyond this point, we have already noted that no other observed states
have a confirmed $P \! = \! -$ eigenvalue, excepting the $Y(4626)$
(which we have argued to be a $c\bar c s\bar s$ state) and $Y(4660)$
(which, in Case~\ref{JunkFit18B}, is a $2P$ state).  The lowest
predicted mass in Table~\ref{tab:AllMasses} and
Fig.~\ref{fig:LevelDiagram} is that of the $I \! = \! 0$, $0^{--}$
state at 4200~MeV, for which $X(4160)$ (currently possessing a very
large mass uncertainty and completely unknown quantum numbers) is at
least a plausible candidate.  Outside of the $1^{--}$ sector, one
finds that all states of $I \! = \! 1$, $J^{PC} \! = \! (0,1,2)^{-+}$
lie between 4210 and 4450~MeV, an enormous range that amusingly almost
exactly matches the mass range of the unconfirmed, $C \! = \! +$
$Z_c(4250)$; very possibly, the $Z_c(4250)$ observation could amount
to the overlapping effect of several of these states.  Lastly, the
unconfirmed $I \! = \! 0$ $X(4350)$ has been suggested as a $c\bar c
s\bar s$ state due to its $J/\psi \, \phi$ decay
mode~\cite{Lebed:2016yvr}, but its mass and $C \! = \! +$ eigenvalue
also make it a candidate $2^{-+}$ state.
\clearpage

\makeatletter\onecolumngrid@push\makeatother

\makeatletter\onecolumngrid@pop\makeatother

\section{Conclusions}
\label{sec:Concl}

In this paper we have studied the lightest hidden-charm
negative-parity exotic meson candidates.  In diquark models
(specifically, the dynamical diquark model) in which these states
consist of two separated diquarks $(cq)(\bar c \bar q')$, one may
classify the states according to Born-Oppenheimer approximation
quantum numbers.  Then the lowest multiplets (as indicated by lattice
QCD simulations of the glue field connecting the diquark pair) are
$\Sigma^+_g(1S)$ and $\Sigma^+_g(1P)$, and all states within these
multiplets carry parity $+$ and $-$, respectively.  The 12
isomultiplets of $\Sigma^+_g(1S)$ were studied previously by
establishing a 3-parameter Hamiltonian that respected all known
phenomenology of the low-lying $P \! = \! +$ states such as $X(3872)$.
The current paper extends this analysis to the 28 isomultiplets of
$\Sigma^+_g(1P)$ by introducing 2 new parameters into the Hamiltonian,
which represent spin-orbit and isospin-dependent tensor terms.

The current experimental status of these $P \! = \! -$ states remains
in flux, notably in the $J^{PC} \! = \! 1^{--}$ sector.  For example,
even the famous $Y(4260)$---known for 15 years---might actually be a
combination of other states like $Y(4230)$, or even a charmonium
hybrid.  Consequently, we have examined a variety of hypotheses for
identifying the 4 $I \! = \! 0$, $J^{PC} \! = \! 1^{--}$ diquark-model
$1P$ states with observed resonances.  An additional and rather
restrictive constraint appears upon noticing that only 1 linear
combination of these 4 states couples to charmonium in a heavy-quark
spin singlet ({\it i.e.}, $h_c$ rather than $\psi$ or $\chi_c$), while
at least 2 of the observed states [$Y(4230)$ and $Y(4390)$] have
significant decays to $h_c$.  Nevertheless, scenarios can be found in
which all current phenomenological constraints are satisfied in a
detailed fit.  In this regard, we have discussed which particular
pieces of data are the most significant ones in allowing or impeding
successful fits.

One remarkable result of the fits is the robustness of the prediction
of the $1P$ multiplet average mass ($\approx \! 4358$~MeV) using only
data from the $1^{--}$ sector, which agrees extremely well with the
result of combining the $1S$ value obtained from previous work with a
calculation of the $1S$-$1P$ splitting using the aforementioned
lattice glue simulations.  The dynamical diquark model appears to be
fully self-consistent in this important regard.

Once the 5 parameters of the Hamiltonian are determined numerically,
then the entire spectrum of 28 isomultiplets is predicted.  Among the
results obtained, we note that the masses of the known $I \! = \! 0$,
$1^{--}$ $Y$ states can easily be fitted in a variety of scenarios.
Somewhat more discriminating is the constraint that the $Y(4230)$ and
$Y(4390)$ have both been observed to decay to $h_c$, but excellent
fits satisfying this additional criterion have also been achieved.
Supposing that $Y(4660)$ is a $2P$ state, the mass prediction of the
sole $I \! = \!  1$, $0^{--}$ state matches that of the observed state
$Z_c(4240)$ that has these quantum numbers.

Future experiments will undoubtedly resolve the ambiguities of
spectroscopy and decay patterns discussed here, making a comparison of
the model to data much more straightforward.  New states may emerge
and old ones may be de-established, or be identified as hybrids rather
than as 4-quark states.  Other Hamiltonian operators essential for
providing important splittings may turn out to have been neglected in
this first attempt, or indeed, the diquark model itself may turn out
to have some fatal flaw.  But even in that worst-case scenario, the
operators used here still have physical significance in a generic
4-quark model, and many more 4-quark states are still predicted to
exist and remain to be discovered.

\begin{acknowledgments}
The authors are grateful to R.E.~Mitchell for important insights into
the current status of the $1^{--}$ sector.  This work was supported by
the National Science Foundation (NSF) under Grant No.\ PHY-1803912.
\end{acknowledgments}

\vspace{2em}

\appendix

\section{Evaluation of the Tensor Operator}
\label{sec:Tensor}

The computation of matrix elements of the tensor operator
\begin{equation}
\label{eq:TensorApp}
S_{12} \equiv 3 \, {\bm \sigma}_1 \! \cdot {\bm r} \, {\bm \sigma}_2
\! \cdot {\bm r} / r^2 - {\bm \sigma}_1 \! \cdot {\bm \sigma}_2 \, ,
\end{equation}
where ${\bm \sigma}$ here denotes not just spin-$\frac 1 2$ Pauli
matrices, but more generally twice the canonically normalized spin
generators ${\bf s}$ for arbitrary spin $s$, is given in many
references, {\it e.g.}, Ref.~\cite{Shalit:1963nuclear}:

\begin{widetext}
\begin{eqnarray} \label{eq:S12general}
\left< L',S',J \right| S_{12} \left| L,S,J \right> & = &
(-1)^{S+J} \sqrt{30[L][L'][S][S']}
\left\{ \begin{array}{ccc}
J & S' & L' \\
2 & L & S
\end{array} \right\}
\left\{ \begin{array}{ccc}
L' & 2 & L \\
0 & 0 & 0
\end{array} \right\}
\left\{ \begin{array}{ccc}
s_1 & s_2 & S \\
s_3 & s_4 & S'\\
1 & 1 & 2 
\end{array} \right\} \left< s_1 || \bm{\sigma}_1 || s_3 \right>
          \left< s_2 || \bm{\sigma}_2 || s_4 \right> \, , \nonumber \\
\end{eqnarray}
\end{widetext}
where again, $[j] \! \equiv \! 2j \! + \! 1$.  The reduced matrix
elements of the angular momentum generators are given by
\begin{equation} \label{eq:Jreduced}
\left< j^\prime || \, {\bf j} \, || \, j \right> = \sqrt{j(2j+1)(j+1)}
\, \delta_{j^\prime j} \, .
\end{equation}
Using Eq.~(\ref{eq:Jreduced}) to simplify Eq.~(\ref{eq:S12general})
and applying the many restrictions on allowed values of angular
momentum that follow, one obtains Eqs.~(\ref{eq:S12simple}).

Section~\ref{sec:MassOps} tabulates the matrix elements of
$S_{12}^{(\qq)}$, {\it i.e.}, the tensor operator for which the
individual spin operators in Eq.~(\ref{eq:TensorApp}) and the total
spins $S, S'$ in Eq.~(\ref{eq:S12general}) refer to the light-quark
pair $\qq$ alone.  In this Appendix we present the results for
$S_{12}^{(\de \bde)}$, the tensor operator for which the individual
spin operators in Eq.~(\ref{eq:TensorApp}) and the total spins $S, S'$
in Eq.~(\ref{eq:S12general}) refer to the diquark pair $\de, \bde$.
The calculation of $S_{12}^{(\de \bde)}$ matrix elements is actually
somewhat simpler than the matrix elements of $S_{12}^{(\qq)}$, since
the total diquark spin $s_{\de \bde}$ to be used in
Eq.~(\ref{eq:S12general}) as $S$ or $S'$ is just the total quark spin
$S$, which is a good quantum number for the $P$-wave states listed in
Table~\ref{tab:PwaveStates}.  Thus, a recoupling such as that in
Eq.~(\ref{eq:JqqCoef}) is not needed.

We now tabulate matrix elements $\Delta M$ of the diquark
spin-dependent operators in Eqs.~(\ref{eq:IsospinDiquark}) and
(\ref{eq:TensorHam2}) for the $1P$ states:
\begin{equation}
\Delta H = {\bm \tau}_q \! \cdot \! {\bm \tau}_{\bar q} \left( V_1
{\bm \sigma}_\de \! \cdot \! {\bm \sigma}_\bde + V^\prime_T \,
S_{12}^{(\de \bde)} \right) \, ,
\end{equation}
where the isospin-dependent part ${\bm \tau}_q \! \cdot \! {\bm
\tau}_{\bar q}$ simply provides a factor of $\{ -3, +1 \}$ for
$I \! = \! \{ 0 , 1 \}$, respectively.  The results, collected by
$J^{PC}$ eigenvalues, are given in the same order as appearing in
Table~\ref{tab:PwaveStates} and in
Eqs.~(\ref{M_1_m_m_I_0})--(\ref{M_2_m_p_I_1}).  However, the matrix
elements for the isospin-independent terms of Eq.~(\ref{eq:FullHam})
($M_0$, $\kappa_{qQ}$, $V_{LS}$) remain the same.
\begin{widetext}
\begin{eqnarray}
\Delta M_{1^{--}}^{I=0} &=&
-6V_1
\begin{pmatrix}
-3 & \sqrt{3} & 0 & 0 \\
\sqrt{3} & -1 & 0 & 0 \\
0 & 0 & 0 & 0 \\
0 & 0 & 0 & 2 \\
\end{pmatrix}
-12V'_T
\begin{pmatrix}
0 & 0 & 0 & \sqrt{\frac{3}{5}}\\
0 & 0 & 0 & -\frac{1}{\sqrt{5}} \\
0 & 0 & 0 & 0 \\
\sqrt{\frac{3}{5}} &-\frac{1}{\sqrt{5}}& 0 & -\frac{7}{5} \\
\end{pmatrix}
\, , \nonumber \\
\Delta M_{1^{--}}^{I=1} &=&
2V_1
\begin{pmatrix}
-3 & \sqrt{3} & 0 & 0 \\
\sqrt{3} & -1 & 0 & 0 \\
0 & 0 & 0 & 0 \\
0 & 0 & 0 & 2 \\
\end{pmatrix}
+4V'_T
\begin{pmatrix}
0 & 0 & 0 & \sqrt{\frac{3}{5}}\\
0 & 0 & 0 & -\frac{1}{\sqrt{5}} \\
0 & 0 & 0 & 0 \\
\sqrt{\frac{3}{5}} &-\frac{1}{\sqrt{5}}& 0 & -\frac{7}{5} \\
\end{pmatrix}
\, , \nonumber \\
\Delta M_{0^{--}}^{I=0} &=& 0 \, , \nonumber \\
\Delta M_{0^{--}}^{I=1} &=& 0 \, , \nonumber \\
\Delta M_{2^{--}}^{I=0} &=&
-12 \left( V_1 + \frac{7}{5} V'_T \right)
\begin{pmatrix}
0 & 0 \\
0 & 1 \\
\end{pmatrix}
\, , \nonumber \\
\Delta M_{2^{--}}^{I=1} &=&
4 \left( V_1 + \frac{7}{5} V'_T \right)
\begin{pmatrix}
0 & 0 \\
0 & 1 \\
\end{pmatrix}
\, , \nonumber \\
\Delta M_{3^{--}}^{I=0} &=&
-12 \left( V_1 - \frac{2}{5} V'_T \right) \, , \nonumber \\
\Delta M_{3^{--}}^{I=1} &=&
4 \left( V_1 - \frac{2}{5} V'_T \right) \, , \nonumber \\
\Delta M_{0^{-+}}^{I=0} &=&
6 \left( V_1 + 2V'_T \right)
\begin{pmatrix}
1 & 1 \\
1 & 1 \\
\end{pmatrix}
\, , \nonumber \\
\Delta M_{0^{-+}}^{I=1} &=&
-2 \left( V_1 + 2V'_T \right)
\begin{pmatrix}
1 & 1 \\
1 & 1 \\
\end{pmatrix}
\, , \nonumber \\
\Delta M_{1^{-+}}^{I=0} &=&
6 \left( V_1 - V'_T \right)
\begin{pmatrix}
1 & 1 \\
1 & 1 \\
\end{pmatrix}
\, , \nonumber \\
\Delta M_{1^{-+}}^{I=1} &=&
-2 \left( V_1 - V'_T \right)
\begin{pmatrix}
1 & 1 \\
1 & 1 \\
\end{pmatrix}
\, , \nonumber \\
\Delta M_{2^{-+}}^{I=0} &=&
6 \left( V_1 + \frac{1}{5} V'_T \right)
\begin{pmatrix}
1 & 1 \\
1 & 1 \\
\end{pmatrix}
\, , \nonumber \\
\Delta M_{2^{-+}}^{I=1} &=&
-2 \left( V_1 + \frac{1}{5} V'_T \right)
\begin{pmatrix}
1 & 1 \\
1 & 1 \\
\end{pmatrix}
\, .
\end{eqnarray} 
\end{widetext}

\bibliographystyle{apsrev4-1}
\bibliography{diquark}
\end{document}